\documentstyle[10pt,epsf,t1enc,fancyhdr,epsfig,subfigure,amsmath,amscd,rotating,here,
multirow,fancybox,indentfirst,dp_delphititle]{dp_delphi}
\makeindex
\def\DpPaperGroup{EP}
\def\DpPaperRef{2000-157}
\def\DpDate{1st March 2000}
\def\DpAuthors{DELPHI Collaboration}
\def\DpSubmit{(Accepted by Eur.Phys.J.C )}
\def\DpTitle{{Measurement of the semileptonic \boldmath $b$ branching 
fractions and average $b$ mixing parameter in $Z$ decays}}
\def\DpComment{ }
\def\DpEMail{ }
\begin{document}

%%% put your own definitions here:
\newcommand{\Zbb}{ Z \rightarrow b \bar b}
\newcommand{\micron}{\mbox{$\mu \mathrm{m}$}}
\newcommand{\Zz}{{\ifmmode Z \else $Z$ \fi}}
\newcommand{\rphi}{\mbox{$R\Phi$}}
\newcommand{\qq}{\ifmmode {q\bar{q} }\else {$q\bar{q}$ }\fi}
\newcommand{\bb}{\ifmmode {b\bar{b} }\else {$b\bar{b}$ }\fi}
\newcommand{\cc}{\ifmmode {c\bar{c} }\else {$c\bar{c}$ }\fi}
\newcommand{\uds}{\ifmmode {uds}\else {$uds$}\fi}
\newcommand{\Dstar}{{ D}^*}
\newcommand{\Dstarp}{{ D}^{*+}}
\newcommand{\Dstarm}{{ D}^{*-}}
\newcommand{\Dstarpm}{{ D}^{*\pm}}
\newcommand{\Dzero}{{ D}^0}
\newcommand{\Dplus}{{ D}^+}
\newcommand{\Dzerob}{{\overline{{ D}^0}}}
\newcommand{\Bzero}{{ B}^0}
\newcommand{\Bzerob}{{\overline{{ B}^0}}}
\newcommand{\Bplus}{{ B}^+}
\newcommand{\Bs}{B_s}
\newcommand{\clX}{ c \rightarrow \ell X}
\newcommand{\BrclX}{{\mathrm{BR}}(\cl X)}
\newcommand{\ra}{\ifmmode {\rightarrow}\else {$\rightarrow$}\fi}
\newcommand{\btol}{\ifmmode {b\ra \ell^-}\else {$b\ra \ell^-$}\fi}
\newcommand{\ctol}{\ifmmode {c\ra \ell^+}\else {$c\ra \ell^+$}\fi}
\newcommand{\bctol} {\ifmmode{b\ra c\ra \ell^+}\else {$b\ra c \ra \ell^+$}\fi}
\newcommand{\bcbtol}{\ifmmode {b\ra \bar{c} \ra \ell^-}\else {$b\ra \bar{c} \ra \ell^-$}\fi}
\newcommand{\Brbl}{{\mathrm {BR}}(\btol)}
\newcommand{\Brbcl}{{\mathrm{BR}(\bctol)}}
\newcommand{\Brbcbl}{{\mathrm{BR}(\bcbtol)}}
\newcommand{\Brcl}{{\mathrm{BR}(\ctol)}}
\newcommand{\ci}{\bar{\chi}}
\newcommand {\bD}{ b \rightarrow D }
\newcommand {\bu}{ b \rightarrow u }
\newcommand {\btaul}{b \rightarrow \tau \rightarrow \ell}
\newcommand{\bpsill}{{b \rightarrow J/\psi \rightarrow \ell^+\ell^-}}
\newcommand{\glcc}{{g \rightarrow c \bar c}}
\newcommand{\glbb}{{g \rightarrow b \bar b}}
\newcommand{\xE}{\left x_E\right}
\newcommand{\xEb}{\left x_E({ b})\right}
\newcommand{\xEc}{ x_E({ c})}
\newcommand{\Lb}{{\Lambda_b}}
\newcommand{\Ds}{{D_s}}
\newcommand{\Lc}{{\Lambda_c}}
\newcommand{\cl}{\rm c \rightarrow \ell }
\newcommand{\bl}{\rm b \rightarrow \ell }
\newcommand{\MeV}{\mathrm{MeV}}
\newcommand{\GeV}{\mathrm{GeV}}
\newcommand{\mm}{\mathrm{mm}}
%========================================================================%
\newcommand {\ErrTot} {\mbox{\sf E}_{\mbox{\sf\scriptsize T}}}
% franco:
\newcommand{\XX}{\ifmmode {\chi^{2}}\else {$\chi^{2}$ }\fi}
\newcommand{\ks}{\ifmmode {k^*} \else {$k^*$} \fi}
\newcommand{\lm}{\ifmmode {\lambda_Q} \else {$\lambda_Q$} \fi}
\newcommand {\bi} {\begin{itemize}}
\newcommand {\ei} {\end{itemize}}
\newcommand {\be} {\begin{equation}}
\newcommand {\ee} {\end{equation}}
\newcommand {\ba} {\begin{eqnarray}}
\newcommand {\ea} {\end{eqnarray}}
\newcommand {\bt} {\begin{table}}
\newcommand {\et} {\end{table}}
\newcommand {\bc} {\begin{center}}
\newcommand {\ec} {\end{center}}
\newcommand {\Rb} {\ifmmode {\mathrm{R}_b} \else $\mathrm{R}_b$ \fi }
%========================================================================%

%%%%%%%%%%%%%%%%%%%%%%%%%% They are a problem with Coll.Sty ?
\makeatletter
%\input{dp_system:coll.sty}
% Collapse citation numbers to ranges.  Non-numeric and undefined labels
% are handled.  No sorting is done.  E.g., 1,3,2,3,4,5,foo,1,2,3,?,4,5
% gives 1,3,2-5,foo,1-3,?,4,5
\newcount\@tempcntc
\def\@citex[#1]#2{\if@filesw\immediate\write\@auxout{\string\citation{#2}}\fi
  \@tempcnta\z@\@tempcntb\m@ne\def\@citea{}\@cite{\@for\@citeb:=#2\do
    {\@ifundefined
       {b@\@citeb}{\@citeo\@tempcntb\m@ne\@citea\def\@citea{,}{\bf ?}\@warning
       {Citation `\@citeb' on page \thepage \space undefined}}%
    {\setbox\z@\hbox{\global\@tempcntc0\csname b@\@citeb\endcsname\relax}%
     \ifnum\@tempcntc=\z@ \@citeo\@tempcntb\m@ne
       \@citea\def\@citea{,}\hbox{\csname b@\@citeb\endcsname}%
     \else
      \advance\@tempcntb\@ne
      \ifnum\@tempcntb=\@tempcntc
      \else\advance\@tempcntb\m@ne\@citeo
      \@tempcnta\@tempcntc\@tempcntb\@tempcntc\fi\fi}}\@citeo}{#1}}
\def\@citeo{\ifnum\@tempcnta>\@tempcntb\else\@citea\def\@citea{,}%
  \ifnum\@tempcnta=\@tempcntb\the\@tempcnta\else
   {\advance\@tempcnta\@ne\ifnum\@tempcnta=\@tempcntb \else \def\@citea{--}\fi
    \advance\@tempcnta\m@ne\the\@tempcnta\@citea\the\@tempcntb}\fi\fi}
 
\makeatother
%%%%%%%%%%%%%%%%%%%%%%%%%% ??????????????????????????????????
% Generate the title page
\begin{titlepage}
\pagenumbering{roman}
\CERNpreprint{\DpPaperGroup}{\DpPaperRef} % Reference of the paper
\date{{\small\DpDate}} % Date of the paper
\title{\DpTitle} % Title of the paper
\address{\DpAuthors} % General name of the author(s)
\begin{shortabs} % Start the abstract
\noindent
%   abstract.tex
%
\noindent

The semileptonic branching fractions for primary and cascade $b$ decays
$\Brbl$, $\Brbcl$ and $\Brbcbl$
were measured in hadronic \Zz decays collected by the
DELPHI experiment at LEP.

The sample was enriched in $b$ decays using the lifetime information 
and various techniques were used to separate leptons from direct or 
cascade $b$ decays.

By fitting the momentum spectra of di-leptons in opposite jets,
the average $b$ mixing parameter $\bar{\chi}$ was also extracted.

The following results have been obtained:

\begin{eqnarray*}
%\begin{center}
% \Brbl = (10.70\pm0.08 (stat)\pm0.21( syst)_{+0.44}^{-0.30}(model) )\% \\
% \Brbcl= ( 7.98\pm0.22 (stat)\pm0.21( syst)^{+0.14}_{-0.20}(model) )\% \\
% \Brbcbl= (1.61\pm0.20 (stat)\pm0.17( syst)^{+0.30}_{-0.44}(model) )\% \\
% \ci   = 0.127\pm0.013 (stat)\pm0.005( syst)\pm0.004(model)\\
 \Brbl &=& (10.70\pm0.08 (stat)\pm0.21( syst)_{+0.44}^{-0.30}(model) )\% \\
 \Brbcl&=& ( 7.98\pm0.22 (stat)\pm0.21( syst)^{+0.14}_{-0.20}(model) )\% \\
 \Brbcbl&=& (1.61\pm0.20 (stat)\pm0.17( syst)^{+0.30}_{-0.44}(model) )\% \\
 \ci   &=& 0.127\pm0.013 (stat)\pm0.005( syst)\pm0.004(model)\\
%\end{center}
\end{eqnarray*}                          

\end{shortabs}
\vfill
\begin{center}
\DpSubmit \ \\ % Horrible hack to allow to have DpSubmit empty
\DpComment \ \\
\DpEMail \ \\
\end{center}
\vfill
\clearpage
\headsep 10.0pt
\addtolength{\textheight}{10mm}
\addtolength{\footskip}{-5mm}
\begingroup
% Commands to process the author names
%
\newcommand{\DpName}[2]{\hbox{#1$^{\ref{#2}}$},\hfill}
\newcommand{\DpNameTwo}[3]{\hbox{#1$^{\ref{#2},\ref{#3}}$},\hfill}
\newcommand{\DpNameThree}[4]{\hbox{#1$^{\ref{#2},\ref{#3},\ref{#4}}$},\hfill}
\newskip\Bigfill \Bigfill = 0pt plus 1000fill
\newcommand{\DpNameLast}[2]{\hbox{#1$^{\ref{#2}}$}\hspace{\Bigfill}}
%
%\small
\footnotesize
\noindent
\DpName{P.Abreu}{LIP}
\DpName{W.Adam}{VIENNA}
\DpName{T.Adye}{RAL}
\DpName{P.Adzic}{DEMOKRITOS}
\DpName{I.Ajinenko}{SERPUKHOV}
\DpName{Z.Albrecht}{KARLSRUHE}
\DpName{T.Alderweireld}{AIM}
\DpName{G.D.Alekseev}{JINR}
\DpName{R.Alemany}{VALENCIA}
\DpName{T.Allmendinger}{KARLSRUHE}
% ...
\normalsize
\endgroup
\titlefoot{Department of Physics and Astronomy, Iowa State
     University, Ames IA 50011-3160, USA
    \label{AMES}}
\titlefoot{Physics Department, Univ. Instelling Antwerpen,
     Universiteitsplein 1, B-2610 Antwerpen, Belgium \\
     \indent~~and IIHE, ULB-VUB,
     Pleinlaan 2, B-1050 Brussels, Belgium \\
     \indent~~and Facult\'e des Sciences,
     Univ. de l'Etat Mons, Av. Maistriau 19, B-7000 Mons, Belgium
    \label{AIM}}
\titlefoot{Physics Laboratory, University of Athens, Solonos Str.
     104, GR-10680 Athens, Greece
    \label{ATHENS}}
\titlefoot{Department of Physics, University of Bergen,
     All\'egaten 55, NO-5007 Bergen, Norway
    \label{BERGEN}}
\titlefoot{Dipartimento di Fisica, Universit\`a di Bologna and INFN,
     Via Irnerio 46, IT-40126 Bologna, Italy
    \label{BOLOGNA}}
\titlefoot{Centro Brasileiro de Pesquisas F\'{\i}sicas, rua Xavier Sigaud 150,
     BR-22290 Rio de Janeiro, Brazil \\
     \indent~~and Depto. de F\'{\i}sica, Pont. Univ. Cat\'olica,
     C.P. 38071 BR-22453 Rio de Janeiro, Brazil \\
     \indent~~and Inst. de F\'{\i}sica, Univ. Estadual do Rio de Janeiro,
     rua S\~{a}o Francisco Xavier 524, Rio de Janeiro, Brazil
    \label{BRASIL}}
\titlefoot{Comenius University, Faculty of Mathematics and Physics,
     Mlynska Dolina, SK-84215 Bratislava, Slovakia
    \label{BRATISLAVA}}
\titlefoot{Coll\`ege de France, Lab. de Physique Corpusculaire, IN2P3-CNRS,
     FR-75231 Paris Cedex 05, France
    \label{CDF}}
\titlefoot{CERN, CH-1211 Geneva 23, Switzerland
    \label{CERN}}
\titlefoot{Institut de Recherches Subatomiques, IN2P3 - CNRS/ULP - BP20,
     FR-67037 Strasbourg Cedex, France
    \label{CRN}}
\titlefoot{Now at DESY-Zeuthen, Platanenallee 6, D-15735 Zeuthen, Germany
    \label{DESY}}
\titlefoot{Institute of Nuclear Physics, N.C.S.R. Demokritos,
     P.O. Box 60228, GR-15310 Athens, Greece
    \label{DEMOKRITOS}}
\titlefoot{FZU, Inst. of Phys. of the C.A.S. High Energy Physics Division,
     Na Slovance 2, CZ-180 40, Praha 8, Czech Republic
    \label{FZU}}
\titlefoot{Dipartimento di Fisica, Universit\`a di Genova and INFN,
     Via Dodecaneso 33, IT-16146 Genova, Italy
    \label{GENOVA}}
\titlefoot{Institut des Sciences Nucl\'eaires, IN2P3-CNRS, Universit\'e
     de Grenoble 1, FR-38026 Grenoble Cedex, France
    \label{GRENOBLE}}
\titlefoot{Helsinki Institute of Physics, HIP,
     P.O. Box 9, FI-00014 Helsinki, Finland
    \label{HELSINKI}}
\titlefoot{Joint Institute for Nuclear Research, Dubna, Head Post
     Office, P.O. Box 79, RU-101 000 Moscow, Russian Federation
    \label{JINR}}
\titlefoot{Institut f\"ur Experimentelle Kernphysik,
     Universit\"at Karlsruhe, Postfach 6980, DE-76128 Karlsruhe,
     Germany
    \label{KARLSRUHE}}
\titlefoot{Institute of Nuclear Physics and University of Mining and Metalurgy,
     Ul. Kawiory 26a, PL-30055 Krakow, Poland
    \label{KRAKOW}}
\titlefoot{Universit\'e de Paris-Sud, Lab. de l'Acc\'el\'erateur
     Lin\'eaire, IN2P3-CNRS, B\^{a}t. 200, FR-91405 Orsay Cedex, France
    \label{LAL}}
\titlefoot{LIP, IST, FCUL - Av. Elias Garcia, 14-$1^{o}$,
     PT-1000 Lisboa Codex, Portugal
    \label{LIP}}
\titlefoot{Department of Physics, University of Liverpool, P.O.
     Box 147, Liverpool L69 3BX, UK
    \label{LIVERPOOL}}
\titlefoot{LPNHE, IN2P3-CNRS, Univ.~Paris VI et VII, Tour 33 (RdC),
     4 place Jussieu, FR-75252 Paris Cedex 05, France
    \label{LPNHE}}
\titlefoot{Department of Physics, University of Lund,
     S\"olvegatan 14, SE-223 63 Lund, Sweden
    \label{LUND}}
\titlefoot{Universit\'e Claude Bernard de Lyon, IPNL, IN2P3-CNRS,
     FR-69622 Villeurbanne Cedex, France
    \label{LYON}}
\titlefoot{Univ. d'Aix - Marseille II - CPP, IN2P3-CNRS,
     FR-13288 Marseille Cedex 09, France
    \label{MARSEILLE}}
\titlefoot{Dipartimento di Fisica, Universit\`a di Milano and INFN-MILANO,
     Via Celoria 16, IT-20133 Milan, Italy
    \label{MILANO}}
\titlefoot{Dipartimento di Fisica, Univ. di Milano-Bicocca and
     INFN-MILANO, Piazza della Scienza 3, IT-20126 Milan, Italy
    \label{MILANO2}}
\titlefoot{Niels Bohr Institute, Blegdamsvej 17,
     DK-2100 Copenhagen {\O}, Denmark
    \label{NBI}}
\titlefoot{IPNP of MFF, Charles Univ., Areal MFF,
     V Holesovickach 2, CZ-180 00, Praha 8, Czech Republic
    \label{NC}}
\titlefoot{NIKHEF, Postbus 41882, NL-1009 DB
     Amsterdam, The Netherlands
    \label{NIKHEF}}
\titlefoot{National Technical University, Physics Department,
     Zografou Campus, GR-15773 Athens, Greece
    \label{NTU-ATHENS}}
\titlefoot{Physics Department, University of Oslo, Blindern,
     NO-1000 Oslo 3, Norway
    \label{OSLO}}
\titlefoot{Dpto. Fisica, Univ. Oviedo, Avda. Calvo Sotelo
     s/n, ES-33007 Oviedo, Spain
    \label{OVIEDO}}
\titlefoot{Department of Physics, University of Oxford,
     Keble Road, Oxford OX1 3RH, UK
    \label{OXFORD}}
\titlefoot{Dipartimento di Fisica, Universit\`a di Padova and
     INFN, Via Marzolo 8, IT-35131 Padua, Italy
    \label{PADOVA}}
\titlefoot{Rutherford Appleton Laboratory, Chilton, Didcot
     OX11 OQX, UK
    \label{RAL}}
\titlefoot{Dipartimento di Fisica, Universit\`a di Roma II and
     INFN, Tor Vergata, IT-00173 Rome, Italy
    \label{ROMA2}}
\titlefoot{Dipartimento di Fisica, Universit\`a di Roma III and
     INFN, Via della Vasca Navale 84, IT-00146 Rome, Italy
    \label{ROMA3}}
\titlefoot{DAPNIA/Service de Physique des Particules,
     CEA-Saclay, FR-91191 Gif-sur-Yvette Cedex, France
    \label{SACLAY}}
\titlefoot{Instituto de Fisica de Cantabria (CSIC-UC), Avda.
     los Castros s/n, ES-39006 Santander, Spain
    \label{SANTANDER}}
\titlefoot{Dipartimento di Fisica, Universit\`a degli Studi di Roma
     La Sapienza, Piazzale Aldo Moro 2, IT-00185 Rome, Italy
    \label{SAPIENZA}}
\titlefoot{Inst. for High Energy Physics, Serpukov
     P.O. Box 35, Protvino, (Moscow Region), Russian Federation
    \label{SERPUKHOV}}
\titlefoot{J. Stefan Institute, Jamova 39, SI-1000 Ljubljana, Slovenia
     and Laboratory for Astroparticle Physics,\\
     \indent~~Nova Gorica Polytechnic, Kostanjeviska 16a, SI-5000 Nova Gorica, Slovenia, \\
     \indent~~and Department of Physics, University of Ljubljana,
     SI-1000 Ljubljana, Slovenia
    \label{SLOVENIJA}}
\titlefoot{Fysikum, Stockholm University,
     Box 6730, SE-113 85 Stockholm, Sweden
    \label{STOCKHOLM}}
\titlefoot{Dipartimento di Fisica Sperimentale, Universit\`a di
     Torino and INFN, Via P. Giuria 1, IT-10125 Turin, Italy
    \label{TORINO}}
\titlefoot{Dipartimento di Fisica, Universit\`a di Trieste and
     INFN, Via A. Valerio 2, IT-34127 Trieste, Italy \\
     \indent~~and Istituto di Fisica, Universit\`a di Udine,
     IT-33100 Udine, Italy
    \label{TU}}
\titlefoot{Univ. Federal do Rio de Janeiro, C.P. 68528
     Cidade Univ., Ilha do Fund\~ao
     BR-21945-970 Rio de Janeiro, Brazil
    \label{UFRJ}}
\titlefoot{Department of Radiation Sciences, University of
     Uppsala, P.O. Box 535, SE-751 21 Uppsala, Sweden
    \label{UPPSALA}}
\titlefoot{IFIC, Valencia-CSIC, and D.F.A.M.N., U. de Valencia,
     Avda. Dr. Moliner 50, ES-46100 Burjassot (Valencia), Spain
    \label{VALENCIA}}
\titlefoot{Institut f\"ur Hochenergiephysik, \"Osterr. Akad.
     d. Wissensch., Nikolsdorfergasse 18, AT-1050 Vienna, Austria
    \label{VIENNA}}
\titlefoot{Inst. Nuclear Studies and University of Warsaw, Ul.
     Hoza 69, PL-00681 Warsaw, Poland
    \label{WARSZAWA}}
\titlefoot{Fachbereich Physik, University of Wuppertal, Postfach
     100 127, DE-42097 Wuppertal, Germany
    \label{WUPPERTAL}}
\addtolength{\textheight}{-10mm}
\addtolength{\footskip}{5mm}
\clearpage
\headsep 30.0pt
\end{titlepage}
%%%%%%%%%%%%%%%%%%%%%%%%%
%
% Change for the document body
%%\pagestyle{heading} % for page numbering
\pagenumbering{arabic} % page numbering in number
\setcounter{footnote}{0} %
\large
%\linenumbers %%%CD
%========================================================================% 

\section{Introduction}

Measurements of the direct semileptonic branching fractions of $b$-hadrons 
are important in order to understand the dynamics of heavy quark decays
and to determine the weak couplings of quarks to the $W$ boson.
From a precise measurement of the inclusive semileptonic branching fractions 
of $b$ quarks a precise value of the Cabibbo-Kobayaski-Maskawa matrix 
element $|V_{cb}|$ can be calculated ~\cite{LEPHS}.

These measurements  have been performed both at the $\Upsilon(4S)$
and in hadronic \Zz  decays. In order to make a comparison between the 
two sets of results, the fact that
the composition of the inclusive sample is different in the two cases
must be taken into account.
At low energy only $B^-$  and $\bar{B^0}$ mesons are produced, while at the 
\Zz, $\bar{B^0_s}$ mesons and $b$-baryons are also present.
Assuming the semileptonic widths of all $b$-hadrons to be equal, their 
respective semileptonic branching fractions are expected to be proportional 
to their measured lifetimes. 
The ratio between the $B^-$  and $\bar{B^0}$ lifetimes to the inclusive 
$b$-hadron lifetime measured at the \Zz, is at present larger than 1,
%($\tau_B / \tau_b = 1.026 \pm 0.020$ ?? \cite{ref:PDG})
 whereas the semileptonic branching fractions of $b$-hadrons 
measured  at the \Zz are slightly larger than the ones measured at the 
$\Upsilon(4S)$ ~\cite{ref:PDG},\cite{Brnew} .
%Recent measurements of the semileptonic branching fractions from the 
%OPAL and L3 Collaborations at LEP give  lower values than first results
% ~\cite{Brnew}, but some discrepancy with respect to measurements at 
%lower energy is still present.

Theoretical calculations which include higher order perturbative QCD 
corrections give a prediction of the branching fraction value 
correlated with the prediction for $<n_c>$, the average number of charmed 
hadrons produced per $b$-hadron decay ~\cite{Neub}.
%, thus introducing another 
%element in the comparison with experimental data.
These results are compatible with the present LEP measurements.
 
In this paper, the two cascade processes:  $\bctol$ and $\bcbtol$ are 
also considered, 
not only because they are the main source of background to the direct
decays, but also  because the values of these branching
fractions  are important inputs to several other heavy flavour measurements,
like asymmetries and oscillations measurements.
The  $\Brbcbl$ measurement
presented in this paper 
is the first inclusive measurement of ``right sign'' leptons 
from cascade decays of $b$-hadrons.
  
In addition, the average $B^0-{\bar{B^0}}$ mixing parameter is measured.
It is the time integrated probability that a $b$-hadron oscillates into
a $\bar b$-hadron:
$\bar{\chi}={ {b \rightarrow {\bar{B^0}}\rightarrow {B^0}\rightarrow 
\ell^+ X} \over { b \rightarrow \ell^{\pm} X}}$. It is related to the mixing
parameters of $B^0_d$ and $B^0_s$ mesons, $\chi_d$ and $\chi_s$ respectively,
by: $\bar{\chi}= g_{B^0_d} \chi_d + g_{B^0_s} \chi_s$, where $g_{B^0_d}$ and 
 $g_{B^0_s}$ are the production fractions of $B^0_d$ and $B^0_s$ in 
semileptonic decays.
Its measurement can therefore be used in the evaluation of the production 
fraction of  $B^0_s$ mesons \cite{LEPHS}.

This paper presents the measurement of inclusive semileptonic branching 
fractions of $b$ quarks in hadronic \Zz  decays using data collected with the 
DELPHI detector at LEP. Four analyses have been 
performed, using different strategies and using various data samples,
partially overlapping.
Events containing $b$ hadrons were selected using lifetime information, 
electrons and muons were identified and several different techniques were 
used to determine the origin of the lepton.
Direct and cascade branching fractions: $\Brbl$, $\Brbcl$ and $\Brbcbl$ were 
measured and, by fitting the momentum spectra of di-leptons in opposite jets,
the average $B^0-{\bar{B^0}}$ mixing parameter $\bar{\chi}$ was also extracted.

The previous DELPHI results on the semileptonic branching fractions 
 ~\cite{ref:gam92} were obtained with data collected at LEP in 1991 and 1992,
using electrons and muons in a sample of hadronic \Zz  decays, with natural 
composition of quark flavours.
A global fit to several electroweak parameters was performed.
With respect to that analysis there is
little dependence on the partial decay widths of the \Zz into \bb and \cc 
quark pairs ($R_b = \Gamma_{b\bar{b}}/\Gamma_{had}$, 
$R_c = \Gamma_{c\bar{c}}/\Gamma_{had}$) 
and the background due to misidentified hadrons and 
leptons from decays and punch-through of light hadrons has been reduced.
The present result supersedes  the previous result obtained by 
DELPHI ~\cite{ref:gam92}.

The layout of the paper is the following:
a description of the DELPHI detector is given in Section~\ref{sec:detector}.  
The selection of the hadronic event sample is described in 
Section~\ref{sec:eventsel}. 
The $b$-flavour tagging algorithm is described in Section~\ref{sec:btag}.
A brief summary of the performances of lepton identification 
algorithms is given in Section~\ref{sec:leptid}. 
Results obtained in the different analyses are then described in the 
following Sections: single and di-lepton analysis (Section~\ref{sec:anal_I}),
single lepton and jet-charge analysis (Section~\ref{sec:anal_II}),
multitag analysis (Section~\ref{sec:anal_III}) and 
inclusive $b$-hadron reconstruction analysis (Section~\ref{sec:anal_IV}).
Finally, in Section~\ref{sec:combi} averages of the results obtained in the 
different analyses are calculated.

\section{The DELPHI detector}
\label{sec:detector}
 
The DELPHI detector has been described in detail in reference~\cite{delphidet}.
Only the components relevant to this analysis are mentioned here.
 
In the barrel region,
the charged particles are measured
by a set of cylindrical tracking detectors with a common axis 
parallel to the 1.2~T solenoidal magnetic field
and to the beam direction.
The time projection chamber (TPC) is the main tracking device.
The TPC is a cylinder with a length of 3 m, an inner radius of
30 cm and an outer radius of 122 cm. 
Tracks are reconstructed using up to 16 space points in the region
$39^\circ < \theta < 141^\circ$,
where $ \theta$ is the polar angle with respect to the beam direction.
Tracks can be reconstructed using at least 4 space points
down to $21^\circ$ and  $159^\circ$.

Additional precise \rphi~measurements,
in the plane perpendicular to the magnetic field,
are provided at larger and smaller radii by the Outer and Inner
detectors, respectively.
The Outer Detector (OD) has five layers of drift cells
at radii between 198 and 206~cm
and covers polar angles from 42$^{\circ}$ to 138$^{\circ}$.
The Inner Detector (ID) is a cylindrical drift chamber
having inner radius of 12~cm and outer radius of 28~cm
and covers polar angles from 23$^{\circ}$ to 157$^{\circ}$.
It contains a jet chamber section providing 24 \rphi~coordinates measurements
surrounded by five layers of proportional chambers with 
both \rphi\ and longitudinal $z$ coordinates measurements.
 
The micro-vertex detector (VD) ~\cite{vdpaper} is located between the LEP
beam pipe and the ID.
It consists of three concentric layers of silicon micro-vertex detectors 
placed at radii of 6.3, 9.0 and 10.9~cm
from the interaction region, called closer, inner and outer layer, 
respectively. 
For all layers the micro-vertex detectors provide hits in the $R\Phi$-plane
with a measured intrinsic resolution of about 8~$\micron$; 
the inner and outer layers provide in addition measurements 
in the $z$~direction, 
with a precision depending on the polar angle and reaching a value 
of~9~$\micron$ for tracks perpendicular to the modules. 
The polar angle coverage for charged particles hitting all three layers of the 
detector is 44$^\circ<\theta< 136^\circ$; the closer layer coverage goes 
down to~25$^\circ$. 
The $z$ measurement was only available in 1994 and 1995.

Additional information for particle identification is provided by the 
Ring Imaging Cherenkov counters (RICH) measuring the Cherenkov light
emitted by particles traversing a dielectric medium faster than
the speed of light. The barrel part of the detector 
covers polar angles from 40$^{\circ}$ to 140$^{\circ}$.
To cover a large momentum range, a liquid ($C_6 F_{14}$) and a gas
($C_5 F_{12}$) radiator are used.
  
The barrel electromagnetic calorimeter, HPC, covers the polar angles between 
$42^\circ$ and $138^\circ$.
It is a gas-sampling device which provides complete three dimensional charge 
information in the same way as a
time projection chamber. Each shower is sampled nine times in its 
longitudinal development. Along the drift direction,
parallel to the DELPHI magnetic field, the shower is sampled every 3.5~mm~; 
in the plane perpendicular to the drift the charge is collected by
cathode pads of variable size, ranging from 2.3~cm in the inner part of the 
detector to 7~cm in the outer layers. 

In the forward regions the tracking is completed by two sets of planar drift 
chambers
(FCA and FCB) placed at distances of $\pm 165$ cm and $\pm 275$ cm from the
interaction point. A lead glass calorimeter (EMF) is used to reconstruct
electromagnetic energy in the forward region.

For the identification of hadronic showers, the iron return yoke 
of the magnet is instrumented with limited streamer mode
detectors to create a sampling gas calorimeter, the Hadronic Calorimeter (HAC).

Muon identification in the barrel region is based on a set of muon
chambers (MUB), covering polar angles between 53$^{\circ}$ and 127$^{\circ}$.
It consists of six active planes of drift chambers,
two inside the return yoke of the magnet after 90~cm of iron
(inner layer) and four outside after a further 20~cm of iron
(outer and peripheral layers).
The inner and outer modules have similar azimuthal coverage.
The gaps in azimuth between adjacent modules are covered by the peripheral
modules. Therefore a muon traverses typically either two inner layer chambers
and two outer layer chambers, or just two peripheral layer chambers.
Each chamber measures the \rphi\ coordinate with a precision of about 2-3~mm.
Measuring \rphi\ in both the inner layer and the outer or peripheral layer
determines the azimuthal angle of muon candidates leaving the return
yoke within about $\pm1^\circ$.
These errors are much smaller than the effects
of multiple scattering on muons traversing the iron.
 
In the forward region the muon identification is done using two sets of planar 
drift chambers (MUF) covering the angular region between $11^\circ$ and 
$45^\circ$. The first set is placed behind 85~cm of iron and the
second one behind an additional 20~cm. Each set consists of two orthogonal layers
of drift chambers where the anode is read out directly and the cathode via a
delay line to measure the coordinate along the wire. The resolution in both
coordinates is about 4~mm.

\section{Event selection}
\label{sec:eventsel} 
%---------------------------------------------------------------------------

Charged particles were accepted if their polar angle was between $20^\circ$
and  $160^\circ$, their track length was larger than $30$ cm,
their impact parameter relative to the interaction point was less than 
5 cm in the plane perpendicular to the beam direction and less than
10 cm along the beam direction and their momentum was larger than 
200 $\MeV/$$c$ with a relative error smaller than 100\%.
Neutral particles detected in the HPC and EMF or in the hadronic calorimeters 
were required to have a measured energy larger than 500 MeV.

The decays of the $Z$ to hadrons were selected by requiring 
a total energy of the charged particles (assumed to be pions) larger 
than 15$\%$ of the center-of-mass energy and at least 7 reconstructed 
charged particles.
With these criteria, the efficiency to select
$q\bar{q}$ events from the simulation was about $95\%$.
All sources of background have been found to 
be below 0.1\%. No significant differences in the acceptance between different 
flavours have been found.

For each event the thrust axis was calculated from the selected charged and 
neutral particles.
Only events with: $|\cos\theta_{thrust}| < 0.90$
were used.
Requiring, in addition, that all sub-detectors needed for these analyses were 
fully operating, totals of about $1 \thinspace 030 \thinspace 000$ and 
$515 \thinspace 000$ \Zz hadronic decays were 
selected from the 1994 and 1995 data samples, respectively.
About $3 \thinspace 800 \thinspace 000$ events were selected from a 
simulated sample of $Z \rightarrow q \bar{q}$ events.
A reduced angular region was used in some parts of the following analyses
to ensure an efficient acceptance for the vertex detector.

Events were generated with the JETSET 7.3 generator
\cite{ref:JETSET} using parton shower and string fragmentation with
parameters optimized to describe the hadronic distributions as
measured by DELPHI \cite{ref:tune}. 
Generated events were passed through a detailed simulation 
%the program DELSIM \cite{ref:DELSIM}
\cite{delphidet} which modeled the detector response and processed through
the same analysis chain as the real data .
Jets were formed from the charged and neutral particles using 
the JADE algorithm with $Y^{min}_{cut} = 0.02$ ~\cite{jade}.
The transverse  momentum of the lepton ( $p_t$ ) was
determined relative to the direction of the jet,
excluding the lepton itself.

Any differences with respect to these selection criteria, as well as their
effect on the statistics used, will be explicitly described for each analysis.
The four analyses used different data subsamples
corresponding to the optimal operation of the subdetectors relevant to the 
definition of the variables used.
Analysis I and IV used 1994 and 1995 data samples, 
Analysis III used also 1992 and 1993 data, while Analysis II used 1994 only.
The 1992 and 1993 statistics are given in Section~\ref{sec:anal_III}.

\section{b-flavour tagging}
\label{sec:btag}

A $b$-flavour tagging algorithm was used in order to obtain a sample enriched 
in $\Zbb$ events.
Events were divided into two hemispheres, with respect to a plane 
perpendicular to the thrust axis and passing through the beam interaction 
point. The $b$-flavour tagging algorithm was applied separately to 
each hemisphere.
Analyses I and IV used
the  combined  $b$-flavour tagging algorithm described in ~\cite{ref:btag}.
This algorithm combines, in a single variable, several quantities which are
sensitive to the presence of a $b$-hadron.

The main discriminant variable is the probability for all tracks belonging
to the hemisphere to come from the primary vertex, calculated from the
 impact parameters of the tracks
positively signed according to the lifetime convention.
Other variables were defined for hemispheres containing a secondary vertex.
% (hemispheres without reconstructed secondary vertices were not considered). ?
These variables are: the effective mass of the system of particles attached to
the secondary vertex, the rapidity of these tracks with respect to the jet
direction and the fraction of the charged energy of the jet which is 
included in the secondary vertex.
Optimized levels of efficiency and purity were chosen in each analysis.

Analysis II used a $b$-flavour tagging algorithm exploiting only the 
information from the impact parameters of charged particles ~\cite{ref:btag}.
Analysis III used a multivariate method to tag the flavours, as 
described in Section ~\ref{sec:III_btag}.

\section{Lepton sample}
\label{sec:leptid}
\subsection{Muon identification}
\label{sec:muid}

To identify a charged particle with momentum greater than 3 GeV$/c$ as a muon
candidate, its track was extrapolated to each of the 
layers of the muon chambers taking into account multiple scattering in the 
material and the propagation of track reconstruction errors.
A fit was then made  between the track extrapolation and the position and
direction of the hits in the muon chambers.
Ambiguities with muon chamber hits associated to more than one extrapolated
track were resolved by selecting the track with the best fit. The charged
particle was then identified  as a muon if the fit was sufficiently good 
and if hits were found outside the return iron yoke.

To exclude regions with poor geometrical acceptance,
a muon was accepted only if its polar angle, $\theta_{\mu} $, was
within one of the following intervals:
$$     0.03 < |\cos\theta_{\mu}| < 0.62 \ \ {\mathrm {or}} \ \  
       0.68 < |\cos\theta_{\mu}| < 0.95, $$

\noindent
which defined the barrel and the forward regions, respectively.

%         **** (exact for Analysis I and IV only, to be specified ?)

The muon identification efficiency was measured in $Z \rightarrow \mu^+\mu^-$ 
events, in the decays of taus into muons and using muons from two-photon
collisions $\gamma \gamma \rightarrow \mu^+\mu^-$.
A  mean efficiency of $0.82 \pm 0.01$ was found with little dependence on 
the muon momentum and on the track polar angle.
Predictions of the simulation agree with corresponding measurements in data, 
both in absolute value and in the momentum dependence, within a precision of 
%2.0 \% and 2.5\% in the barrel and in the forward region, respectively.
1.5\%.
 
An estimate  of the misidentification probability 
was obtained by means of a lifetime-based anti $b$-tag to select a 
background enriched sample. 
After the subtraction of the muon content in the selected sample,
the misidentification probability was found to be $(0.52\pm0.03)$\% 
in the barrel and $(0.36\pm0.06)$\% in the forward regions.
Applying the same procedure to the simulation gave however lower values,
with factors
%The ratio with the value of the same quantity measured in simulated events 
%was found to be in average
$2.03 \pm 0.12$ $(2.02 \pm 0.13)$ in the barrel and 
$1.22 \pm 0.20$ $(1.78 \pm 0.24)$ in the forward regions for the 
1994 (1995) samples, respectively, showing a small momentum dependence
and about 30\% reduction near the borders of the geometrical acceptance 
of the muon chambers.

The hadron misidentification probability, measured both in data and in 
simulation, was cross-checked using pions from $K_{s}^{0}$ and $\tau$ decays 
and compatible results were found.
In Analysis I, II  and IV the simulated hadrons misidentified as muons were 
reweighted according to the probability measured in data.
In Analysis III a different approach was used to estimate the misidentification
 probability,  as described in Section \ref{sec:III_back},
and good agreement with the above results was found.

\subsection{Electron identification}
\label{sec:elecid}

Charged particles with momenta greater than  3 GeV$/c$ 
and within the efficient acceptance region of the HPC 
($0.03 < |\cos\theta_{e}| < 0.72$)
%       **** (exact for Analysis I and IV only, to be specified ?)
were selected as electron candidates 
on the basis of the information from the HPC, the TPC and the RICH detectors.
Tracks were extrapolated to the HPC and associated to detected showers.
The signals from the various detectors were then analyzed by a neural 
network.
By using the network response obtained in a sample of simulated electrons from 
$b$ and $c$ decays, a momentum dependent cut was defined in order to have a 
65\%~efficiency, constant over the full momentum range. 

To reduce the contamination from electrons produced from photon conversions, 
electron candidates were removed if they came from a secondary vertex 
and carried no transverse momentum relative to the direction from the 
primary to this  secondary vertex.

The efficiency of tagging an electron was measured in the data 
by means of a sample of isolated electrons extracted from selected 
Compton events and a sample of electrons produced from photon conversions 
in the detector. 
The ratio between the values of the efficiencies measured in real and 
simulated events was parameterized in terms 
of the $p_t$ and the polar angle of the track and found to be on average
$0.92 \pm 0.02$ and
$0.93 \pm 0.02$, in the 1994 and 1995 samples, respectively.
A corresponding correction factor was then applied to the sample of electrons 
in  simulated $\qq$ events.
  
The probability of tagging a hadron as an electron was also measured in the 
data by selecting a background sample by means of the  anti $b$-tag technique 
in the same manner as for muons. 
The measured misidentification probability in data and the ratio with the 
same quantity obtained in simulated events were on average 
$(0.40 \pm 0.02)\%$ and $0.76 \pm 0.05$ in the 1994 sample 
and $(0.38 \pm 0.04)\%$ and $0.70 \pm 0.06$ in the 1995 sample.
%, showing a few percent relative increase at the HPC boundary.

\subsection{Simulated lepton sample}
\label{sec:model}

Samples of simulated events, which
were processed through the same analysis chain as the
data as described in Section \ref{sec:eventsel}, were used to obtain
reference spectra for the different sources of simulated leptons.
\par
The $b$ semileptonic decays to electrons and muons were simulated using the 
model of Isgur et al. \cite{ISGW} (ISGW  model in the following).
The model of Bauer et al. \cite{ref:WSB}, which takes into account the 
finite mass of the produced lepton, was used
for the $b$ decays into $\tau$'s. For $D$ decays the branching
ratios were adjusted to be in better agreement with measured values 
~\cite{ref:PDG}. 
In the different semileptonic decay modes,
the branching fractions for the decays to neutral pions, when not measured, 
were obtained imposing isospin invariance. 
Reference spectra with alternative models have been obtained
reweighting the events according to the decay model considered.
The weight was computed on the basis of the lepton momentum in the
$B(D)$ rest frame.
According to the prescription of ~\cite{ref:LEPHF}, 
for the central value of the results, the inclusive model of Altarelli et al.
\cite{ACCMM} (ACCMM model in the following) was used,
with model parameters tuned to the CLEO data ~\cite{CLEO92}, whereas 
ISGW and  ISGW$^{**}$ models have been  used to  evaluate the systematic 
uncertainties. ISGW$^{**}$ indicates the ISGW model modified to 
include a 32\% contribution of charmed excited states 
(referred to as $D^{**}$), instead of the original 11\% predicted by the
model itself, so as to better describe the CLEO data.

Leptons from the decay chain ${ b\rightarrow c W \rightarrow c\bar{c}q 
\rightarrow c \ell^- X}$ (the so called ``upper decay vertex'') were 
considered with the contributions from both $D_s\rightarrow \ell^-X$ and 
$\bar{D^0} (D^-) \rightarrow \ell^-X$.

%*** also for Analysis II and III ? 

%************************************************************************** 
\section{Analysis I: Measurement of semileptonic $b$ decays from
single leptons and di-leptons spectra}
\label{sec:anal_I}

In this analysis the semileptonic branching fractions for primary and cascade 
$b$ decays $\Brbl$,  $\Brbcl$ , $\Brbcbl$ and the average $b$ mixing parameter, 
$\ci$, are measured using the momentum spectra of single lepton and 
di-leptons in opposite jets.
The single lepton spectra are studied in a 
sample of events highly enriched in  $\bb$, selected by means of a $b$-flavour 
tagging algorithm. In the di-lepton sample, the $\bb$ purity is increased  
by requiring a minimum $p_t$ for one of the leptons.

The sensitivity to the different sources of leptons is given by the
kinematic properties of leptons from different sources and by the charge 
correlation between di-leptons in opposite jets from $b$ and $\bar{b}$,
respectively.

Hadronic events and lepton candidates were selected as described in Sections
~\ref{sec:eventsel} and ~\ref{sec:leptid}. The angular region 
$|\cos\theta_{thrust}| < 0.9$ was used for di-lepton candidates, while
for single lepton events, to have a good efficiency in the $b$-flavour 
tagging, 
events were considered only if they fulfilled  $|\cos\theta_{thrust}| < 0.7$. 
As a consequence, only barrel muon chambers were considered for single muons.
About $768 \thinspace 000$ and $385 \thinspace 000$ \Zz hadronic decays 
were selected in the 1994 and 1995 data samples, respectively.

\subsection{Single lepton fit}
\label{sec:I_single}

Events were divided into two hemispheres  with respect to a plane 
perpendicular to the thrust axis and passing through the beam 
interaction point.
A primary vertex was reconstructed in each hemisphere to suppress possible 
correlations between the two hemispheres induced by the $b$-tagging algorithm.
The combined  $b$-flavour tagging algorithm described in 
Section \ref{sec:btag}
 was used to select hemispheres enriched in $b$-hadron content while, 
in the opposite  hemisphere, the single lepton spectra were studied.
For the cut on the combined $b$-tagging variable used in this analysis,
the following efficiencies
for selecting different flavours were estimated from simulation:
$ \varepsilon_b      =  ( 39.34 \pm 0.05 ) $\%, 
$ \varepsilon_c      =  ( 1.87 \pm 0.02 ) $\%,
$ \varepsilon_{uds}  =   ( 0.189 \pm 0.003) $ \%, 
so that the fraction of $b$ events in the sample was ${\cal P}_b = 95.1 \%$. 

The value of $\varepsilon_b$ is quoted only for reference, since it
is never used in the following.
In practice the number $N^H_b$ of tagged hemispheres which contain a $b$
quark was estimated as:
$$N^H_b = N^H_{tag} - (\varepsilon_c \times R_c + \varepsilon_{uds} \times
 R_{uds}) \times 2 N_{had}$$

\noindent
 where: $N^H_{tag}$ and $N_{had}$ are the total numbers of tagged hemispheres  
 and the number of hadronic  events,  respectively,
$\varepsilon_c$ and  $\varepsilon_{uds}$ were the efficiencies 
for charm and light quark events, respectively, obtained from  simulation,
 and $R_{uds}= \Gamma_{uds}/\Gamma_{had}$ = $1 - R_{b} - R_{c} $.  
% were the ratios of partial decay widths to $c$ and $uds$ quarks.
The LEP averages of $0.21643\pm 0.00073$ and $0.1694\pm0.0038$ 
were used for $R_{b}$ and $R_{c}$, respectively ~\cite{ref:LEP99}. 
The number of $b \bar{b}$ events used in the simulation was normalized to 
the same value $N^H_b$.

Once a hemisphere was tagged as $b$, leptons were studied in the opposite
hemisphere. 
A correction was applied, estimated from simulation, 
because of the correlation between the lifetime and the lepton tags.  
It arose  mainly  from the acceptance requirements,
which are different for electrons and muons, and amounted to
 $\rho_e=1.003 \pm 0.005$ and $\rho_\mu=1.017 \pm 0.005$. 
Here $\rho$ is the fraction of lepton candidates found in the hemisphere 
opposite to the $b$-flavour tagged hemisphere, compared to
the fraction of lepton candidates found in an unbiased $b$ hemisphere.
Before calculating the lepton transverse momentum, a search for secondary 
vertices was performed using the same algorithm as in ~\cite{ref:btag}.
%If a secondary vertex was present in the jet, the jet direction was corrected 
%using the primary to secondary vertex direction.   
When the secondary vertex was successfully reconstructed 
(about 45\% of the events), 
the primary to secondary vertex direction was found to give a better 
approximation of the $b$-hadron flight direction than the jet axis, and was 
used in its place. The resolution on the $b$-hadron flight direction improved 
correspondingly  from 30 to 20 mrad.

Lepton candidates were classified according to their different origin
as follows:
\begin{itemize}
\item[a)] direct $b$-decay: \\
           ${ b \rightarrow \ell^- + X }$, 
\item[b)] ``right sign'' cascade decays: \\
            ${b \rightarrow \bar{c} + X \rightarrow \ell^- + X}$,
\item[c)] ``wrong sign'' cascade decays: \\
            ${b \rightarrow c + X \rightarrow \ell^+ + X}$, 
\item[d)] $b$ decays into $\tau$ lepton:\\
            ${b \rightarrow \tau^- + X \rightarrow \ell^- + X}$, 
\item[e)] direct $c$-decay \\
            ${c \rightarrow \ell^+  + X}$, 
%            ${c \rightarrow \tau^+ + X  \rightarrow \ell^+ + X}$,\\
\item[f)] prompt leptons from $J/\Psi$ decays or from $b$ or $c$ decays, 
where the ${c \bar c}$ (${b \bar b}$) pair is produced by 
gluon splitting,
\item[g)] misidentified or decaying hadrons.
\end{itemize}
The above classification was considered both for electrons and muons, 
separately.

A binned maximum likelihood fit was used to compare the
momentum and transverse momentum spectra of electrons and muons in data with
the simulation. The full likelihood  expression is reported in appendix.

\subsection{Di-lepton fit}
\label{sec:I_double}

The single lepton likelihood was multiplied by a likelihood obtained for 
di-leptons in opposite hemispheres, in order to separate the $\btol$ from
the $\bctol$ and the $\bcbtol$ components and to extract the 
average mixing parameter $\ci$. 
In the di-lepton sample no $b$-flavour tag was used  in order
not to introduce any bias in the composition of the $b$-hadron sample.
The $b$ enrichment was obtained by requiring a minimum $p_t$ for one of the two
leptons. The full $p_t$ spectrum was considered for the opposite lepton.
For a cut at $p_t > 1.2$ GeV$/c$, a $b$ purity of about  88\% was obtained  
using simulated events.

Di-lepton events were separated, for  both the data and the simulated samples,
into six groups depending on whether the two  lepton candidates 
have the same or opposite charge and on which combination of lepton
species ($ee, \, e\mu, \, \mu\mu$) they belonged to.
Lepton pairs were used if the two leptons were separated by at least $90^o$,
while lepton pairs coming from the same jet were omitted from the fit
to avoid additional systematic uncertainties in the composition of the
cascade lepton sample.
In each group, simulated events were separated  into di-lepton classes,
 according to the different possible combinations in the two hemispheres
 of the above mentioned single-lepton classes (a) to (g).
To guarantee a reasonable number of events in each bin, 
the $p$ and $p_t$ of each 
lepton in the pair were combined to form a single variable,
the combined momentum, $p_c$, defined as in ~\cite{pcomb}: 
$p_c\,=\,\sqrt{p_t^2+\frac{p^2}{100}}$.
Two-dimensional  reference distributions were obtained for
the chosen combinations in the variables  $(p_c^{min},p_c^{max})$, where 
$p_c^{min}$ ($p_c^{max}$)
refers to the smaller (larger) combined  momentum.

If $B^0-\bar{B^0}$ mixing is not considered, the main source of di-leptons 
having opposite charges are direct $b$-decays: 
${ (b \rightarrow \ell^-) ( \bar b \rightarrow \ell^+) }$. But, in the 
presence of mixing, a fraction $2\ci(1-\ci)$ of these di-leptons
have the same charge.
Same charge di-leptons also arise from events with one 
direct $b$-decay and one cascade $b$-decay: 
${( b \rightarrow \ell^- )(\bar b \rightarrow \bar c \rightarrow \ell^- )}$.
Because of  mixing, a fraction $2\ci(1-\ci)$ of these events will enter the
opposite charge class.

The fraction of leptons of class a, b and c were determined by the fit,
whereas contributions from lepton classes (d) to (g) were fixed to the values
given in Table~\ref{tab:I_sys}.
The detailed expression of the likelihood function, for single lepton  
and di-lepton, is reported in appendix.

\subsection{ Results and systematic uncertainties}
\label{sec:I_res}

The results obtained with the 1994 and 1995 samples and their average are
shown in Table \ref{tab:results}, where the uncertainties 
are statistical only.
About 12\% of the single leptons were also included in the di-lepton sample
and the statistical uncertainties have been corrected accordingly.

\begin{table}[htb]
%\begin{Table}[p]
\begin{center}
\begin{tabular}{|l|c|c|c|}
\hline
             &  1994             &   1995            & 1994+1995 \\
\hline
 $\Brbl $    & $0.1066\pm0.0014$ & $0.1081\pm0.0019$ & $0.1071\pm0.0011$  \\
 $\Brbcl$    & $0.0822\pm0.0049$ & $0.0781\pm0.0064$ & $0.0805\pm0.0039$  \\
 $\Brbcbl$   & $0.0144\pm0.0044$ & $0.0196\pm0.0056$ & $0.0164\pm0.0035$  \\
 $\ci$       & $0.119 \pm0.016$  & $0.138 \pm0.022$  & $0.126\pm0.013$ \\
\hline
\end{tabular}
\end{center}
\caption{Results of the fit to the 1994 and 1995 lepton samples and their
combination. The uncertainties are statistical only.}
\label{tab:results} 
\end{table}

\begin{figure}[p]
\begin{center}
\mbox{\epsfig{file=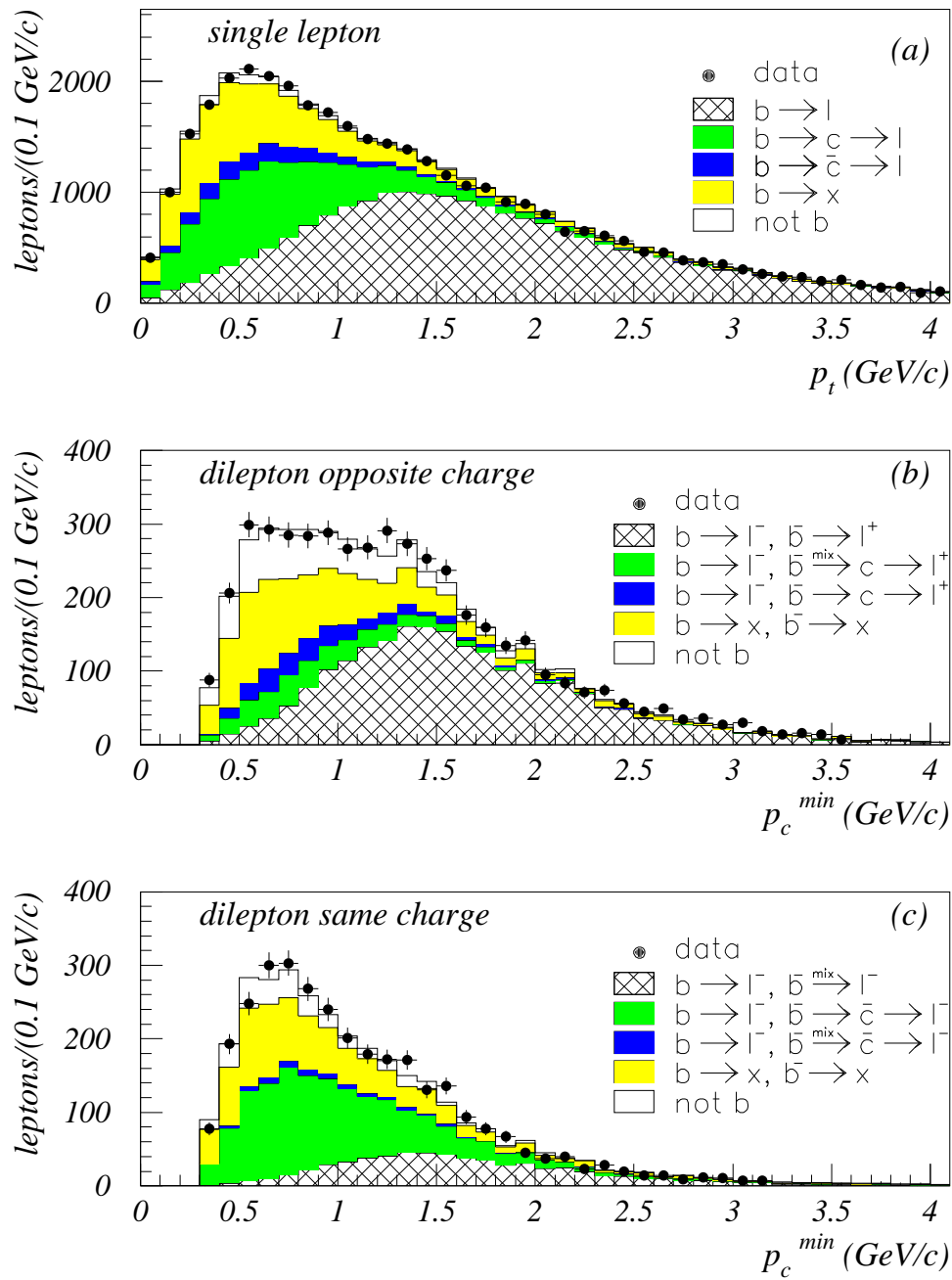,height=20cm}}
\end{center}
\caption{ Comparison of data and simulation spectra.
The simulation spectra have been  reweighted according to the result of the 
fit.
(a) Transverse momentum distribution for single electrons and muons.
$\mathrm {b \rightarrow x}$ indicates $b$ decays to misidentified or 
decaying hadrons.
(b)((c)) Combined momentum distribution for the two leptons in 
di-lepton events, identified in opposite jets and having the 
opposite (same) charge.
$p_c^{min}$ refers to the minimum combined momentum of the two leptons.
In the legend of (b) and (c) the lepton origin in the two hemispheres 
is described, the label ``mix'' refers to events where  
$B^0-\bar{B^0}$ mixing occurred.
 }
\label{fig:btag}
\end{figure}

%The Peterson fragmentation parameter \cite{ref:pet83}, $\epsilon_b$, was 
%left free to vary in the fit.
%Converted into the mean fractional energy of $b$-flavoured hadrons 
%it gives $x_E = 0.7126 \pm 0.0031$, where the uncertainty is statistical only.
In Figure \ref{fig:btag} single lepton and di-lepton spectra are shown.
The simulation spectra have been 
reweighted according to the result of the fit.
The correlation matrix for the statistical uncertainties 
is shown in Table \ref{tab:mate}.

\begin{table}[htb]
%\begin{Table}[p]
\begin{center}
\begin{tabular}{|l|cccc|}
\hline
          &  $\Brbl$ & $\Brbcl$ &$\Brbcbl$&  $\ci$  \\
\hline
 $\Brbl$     & 1.00  & -0.241   & -0.061  &  0.086  \\
 $\Brbcl$    &       &  1.00    & -0.797  & -0.159  \\
 $\Brbcbl$   &       &          &  1.00   &  0.112  \\
 $\ci$       &       &          &         &  1.00   \\
\hline
\end{tabular}
\end{center}
\caption{Correlation matrix of statistical uncertainties in Analysis I.
}
\label{tab:mate} 
\end{table}

\begin{table}
\begin{center}
\begin{sideways}%
\begin{minipage}{\textheight}
\begin{center} 
\begin{tabular}{|l|c|c|c|c|c|}
\hline
Error Source &Range & $\Delta\Brbl$ &$\Delta\Brbcl$&$\Delta\Brbcbl$& $\Delta\ci$  \\ 
             &      & $10^{-2}$     & $10^{-2}$    & $10^{-2}$     & $10^{-2}$ \\ \hline

electron efficiency & $\pm 3 \%$   & $\mp$0.15  & $\mp$0.14 & $\mp$0.06& $\pm$0.02  \\ 
misidentified e      & $\pm 8 \%$   & $\mp$0.05  & $\mp$0.14 & $\mp$0.06& $\pm$0.04  \\ 
converted photons    & $\pm 10 \%$  & $<$0.01    & $\mp$0.06 & $\mp$0.03& $\pm$0.01 \\ 
$\mu$ efficiency     &$\pm2.5\%$  & $\mp$0.14  & $\mp$0.18 & $\mp$0.05& $\pm$0.06  \\ 
misid. $\mu$ barrel, forward &$\pm6.5\% , $17\%&$\mp$0.01  & $\mp$0.15& $\mp$0.06& $\pm$0.02  \\ 
jet direction    & see text      & +0.05      & -0.03     & -0.08    & + 0.6    \\ 
\hline
 $\varepsilon_c$      & $\pm 9 \%$  & $\pm$0.02  & $\mp$0.01 &$\mp$0.01 &$\pm$0.03 \\ 
 $\varepsilon_{uds}$  & $\pm 22\%$  & $\pm$0.01  & $\pm$0.02 &$<$0.01 &$\mp$0.02 \\ 
 $\ell-b$ correlation & $\pm 1\%$  & $\mp$0.05  & $\mp$0.11 &$\mp$0.03 &$\pm$0.03 \\ 
 $\ell-b$ corr. $p$ dependence & see text  &$\mp$ 0.04  &$\pm$ 0.03 & $\mp$0.01 & $\mp$ 0.04 \\ 
 $R_b$        & $0.21643\pm 0.00073$~\cite{ref:LEP99}
                             & $<$0.01    & $<$0.01   & $<$0.01 &$<$0.01   \\ 
 $R_c$        & $0.1694\pm 0.0038$ ~\cite{ref:LEP99} 
                     & $<$0.01  & $<$0.01 & $<$0.01 & $<$0.01  \\ 
\hline
$x_E(b)$        & $0.702 \pm 0.008$~\cite{ref:LEPHF} & $\mp$0.11  & $\pm$0.07 &$\pm$0.04 
&$\mp$0.15 \\
$x_E(c)$        & $0.484 \pm 0.008$~\cite{ref:LEPHF} & $\mp$0.02  & $\pm$0.03 &$\mp$0.03 &$\pm$0.02 \\

${b\rightarrow W\rightarrow D}\over{b\rightarrow W\rightarrow D_s}$
 &$(1.28^{+1.52}_{-0.61})$~\cite{ref:LEPHF} 
       &$\pm$0.03&$^{+0.20}_{-0.11}$ &$^{-0.23}_{+0.13}$ &$^{-0.09}_{+0.07}$ \\

BR($\btaul$)   &$(0.459\pm0.071)$\%~\cite{ref:PDG}  
               & $\mp$0.02 & $\mp$0.03 &$\mp$0.04 &$\pm$0.02 \\
BR($\bpsill$)  &$(0.07\pm 0.01)$\%~\cite{ref:PDG}  
               & $\mp$0.03 & $\pm$0.01 &$\pm$0.01 &$\mp$0.09 \\
$\Brcl$        &$(9.85\pm0.32)$\%~\cite{ref:LEP99}  
               & $\mp$0.01 & $\mp$0.03 &$\mp$0.04 &$\pm$0.01 \\ 
$\glcc$       & $(3.19\pm0.46)$\%~\cite{ref:LEP99}  
              & $<$0.01   & $<$0.01   &$<$0.01&$<$0.01\\
$\glbb$       & $(0.251\pm0.063)$\%~\cite{ref:LEP99}  
              & $<$0.01   & $<$0.01   &$<$0.01&$\pm$0.01\\
\hline
 total systematic & & $\pm$0.26  & $\pm$0.38 &$\pm$0.25 &$\pm$0.64  \\ 
\hline
Semilept.mod.$\bl$\cite{ref:LEPHF}
              & ACCMM ($\mathrm{^{+ISGW}_{-ISGW**}}$)
&$^{-0.24}_{+0.41}$&$^{+0.23}_{-0.29}$&$^{+0.14}_{-0.23}$&$^{-0.23}_{+0.28}$\\

Semilept.mod.$\cl$\cite{ref:LEPHF}
              & ACCMM1($\mathrm{^{+ACCMM2}_{-ACCMM3}}$)  
              & $^{-0.08}_{+0.07}$ & $^{-0.11}_{+0.01}$& $^{-0.03}_{+0.02}$ &$^{-0.33}_{+0.34}$ \\ 

\hline

 total models & &\large{$^{-0.25}_{+0.42}$} &\large{$_{-0.31}^{+0.23}$} &
                 \large{$_{-0.23}^{+0.14}$} &\large{$^{-0.40}_{+0.44}$} \\ 

\hline
\end{tabular}
\end{center}
\caption[bla]{Summary of systematic uncertainties in the 
analysis of single and di-lepton events.
Ranges given  in \% correspond to  relative variations around the
central value.
}
\label{tab:I_sys}
\end{minipage}
\end{sideways}
\end{center}
\end{table}

The following sources of systematic uncertainties have been considered:
\begin{itemize}
\item experimental uncertainty related to lepton measurements: 

the muon and electron identification efficiencies and the background due 
to hadron misidentification 
have been varied considering their measurement uncertainties 
in the data-simulation
comparisons (see Sections \ref{sec:muid},\ref{sec:elecid}).
To account for effects related to the difference in topology between the 
test samples used in Sections \ref{sec:muid},\ref{sec:elecid} and the 
hadronic environment, an additional uncertainty of $\pm$ 2\% has been 
applied to the efficiencies, as estimated from simulation.
As a consequence, the total relative uncertainties assumed on the leptons 
efficiencies were
$\pm$ 2.5\% and  $\pm$ 3\% for muons and electrons, respectively.
The residual contamination in the electron sample due to converted photons
has been varied by $\pm$ 10\%.

The angular distribution between di-leptons is well described by simulation,
therefore the angular cut of $90^o$ is assumed not to add any systematic 
uncertainty.

%The systematic error due to the uncertainty on the $b$-quark direction and
%consequently on the lepton transverse momentum has been evaluated comparing
%the jet momentum direction with the direction determined by the secondary
%vertex in case it was successfully reconstructed. The mean difference in the 
%jet direction was found to be $1.5^\circ$; the
%fit has then been performed using both methods and half difference on the
%results has been used as systematic error.
 
The fit has been performed using for the $p_t$ calculation both the jet
direction and the secondary vertex direction. Half the difference between the
results has been used as systematic uncertainty.

\item experimental uncertainty related to the $b$-flavour tagging:

efficiencies to tag $c$ and $uds$ quarks have been varied by 9\% and 
22\%, respectively, according to the uncertainties in ~\cite{ref:btag}. 
The partial decay widths $R_b$ and $R_c$
have been varied according to their measurement uncertainties.
%The correlation between the lifetime and the lepton tags has been varied 
%by twice its statistical uncertainty.

The correction factors for the correlation between the $b$-tag and the 
leptons ($\rho_e$ , $\rho_\mu$) have been varied 
by twice their statistical uncertainties.
The dependence on lepton momentum of the correlation has also been studied. 
Since the $b$-tag efficiency is higher in presence of high momentum leptons,
the lepton spectrum in hemispheres opposite to a $b$-tagged one is slightly 
biased towards low momenta. A correction has been estimated with simulation
comparing spectra in tagged and non tagged events and the full effect has 
been assumed as a systematic uncertainty.

The stability of the result as a function of the cut on the $b$-flavour tagging
variable has been checked to be compatible with the corresponding 
statistical fluctuations.

\item modelling uncertainty related to the assumed physical parameters:

the mean value and the range of variation of several physical parameters 
used in the simulation was calculated according to references
 \cite{ref:PDG}, \cite{ref:LEPHF} and \cite{ref:LEP99}.
In particular they have been varied: the  mean fractional energy of $b$ and 
$c$ hadrons, the branching fractions assumed  for $b\to\tau\to\ell$, 
$b\to J/\Psi\to\ell$, $c\to\ell$ and the fraction of gluon splitting  
to heavy quarks.
The lepton distribution from the ``upper vertex'' was studied by
varying the contributions of $D_s \to \ell^-X$ and 
$\bar{D}^0(D^-) \to \ell^-X$ of the amount suggested in \cite{ref:LEPHF}.
Varying the B hadron composition was found to produce negligible effect.

\item the modelling uncertainty related to different semileptonic 
decay models has been calculated according to ~\cite{ref:LEPHF}. 
Thus the ISGW and ISGW$^{**}$ models have been used as 
conventional references for evaluating the semileptonic decay model 
uncertainty and this variation represents the dominant source of systematic 
uncertainty.

\item the finite statistics used in the simulation was checked to introduce a 
negligible systematic error.

\end{itemize}
        
The summary of systematic uncertainties is given in Table \ref{tab:I_sys}.

In conclusion 
from a fit to single and di-lepton events from data collected with the DELPHI
detector in 1994 and 1995, the semileptonic branching fractions $\Brbl$,  
$\Brbcl$, $\Brbcbl$ and the average $b$ mixing parameter $\ci$ have been 
measured:
\par
\begin{eqnarray*}
 \Brbl &=& (10.71\pm0.11 (stat)\pm0.26( syst)^{-0.25}_{+0.42}(model) )\% \\
 \Brbcl&=& (8.05 \pm0.39 (stat)\pm0.38( syst)_{-0.31}^{+0.23}(model) )\% \\
 \Brbcbl&=& (1.64\pm0.35 (stat)\pm0.25( syst)_{-0.23}^{+0.14}(model) )\% \\
 \ci   &=&0.126  \pm0.013(stat)\pm0.006( syst)\pm0.004(model)\\
\end{eqnarray*}                          

%**************************************************************************
\section{Analysis II: Measurement of semileptonic $b$ decays from single leptons
 and jet-charge}
\label{sec:anal_II}
In this analysis a
sample of $b$ enriched events was obtained by applying $b$-flavour tagging 
separately to each hemisphere of the event,
 only events with the thrust axis contained in 
the region |$cos\theta_{thrust}$| < 0.8 were used.
The $b$ tagging algorithm exploited only the information from the impact 
parameters of the 
tracks from charged particles assigned to the hemisphere: the cut 
selected 
69.2 \% of \bb, 12.9 \% of \cc\ and 1.1 \% of \uds\
events, so that the fraction of $b$ events in the sample was 
${\cal P}_b = 84.0 \%$. 
Leptons were  selected from all the charged particles with momentum 
$p > 3$  GeV$/c$, lying in the hemisphere
opposite to the $b$-tagged hemisphere within the acceptance of the HPC 
or muon chambers. 

The lepton was then used as a seed to reconstruct the position of the $b$ 
decay vertex, by applying the algorithm
originally developed for lifetime and oscillation measurements (for details,
 see e.g. \cite{blife}). A vertex was
found in 92.5 $\pm$ 0.2 (92.3 $\pm$ 0.1)\% of the cases in the data 
(simulation). The direction of 
the $b$-hadron was then obtained by 
averaging the direction of the jet containing the lepton with the one of the 
vector 
joining the primary to the secondary vertex: when the vertex was not 
reconstructed, only the jet direction was used.
The energy of the $b$ hadron was computed from the sum of the energy of the 
charged and 
neutral particles assigned to its jet and the missing energy in the 
hemisphere (computed as described 
in \cite{vcb}). The resolution was $\sigma (E_B)/E_B \simeq 12 \%$. 
This allowed the entire $b$-hadron four-momentum to be reconstructed, 
by assuming an average mass of $\simeq$ 5.3 GeV/$c^2$.

Leptons from direct \btol decays were then separated from the other sources
of leptons 
%( \bctol, \ctol, fake hadrons, etc.)
by means of kinematics and charge correlation, as described in the following. 
The momentum of the lepton in the $b$-hadron rest frame, $k^*$,
 was computed by boosting back the lepton into the $b$-hadron rest frame: 
the resolution was about $\sigma_\ks \simeq 200$ MeV/$c$. 
The \ks\ spectra for \btol, \bctol, \ctol\ decays in the simulation 
were tuned as described in section \ref{sec:model} and varied according to 
the prescriptions already described to compute the systematic uncertainty. 

\begin{figure}
\bc
\epsfig{file=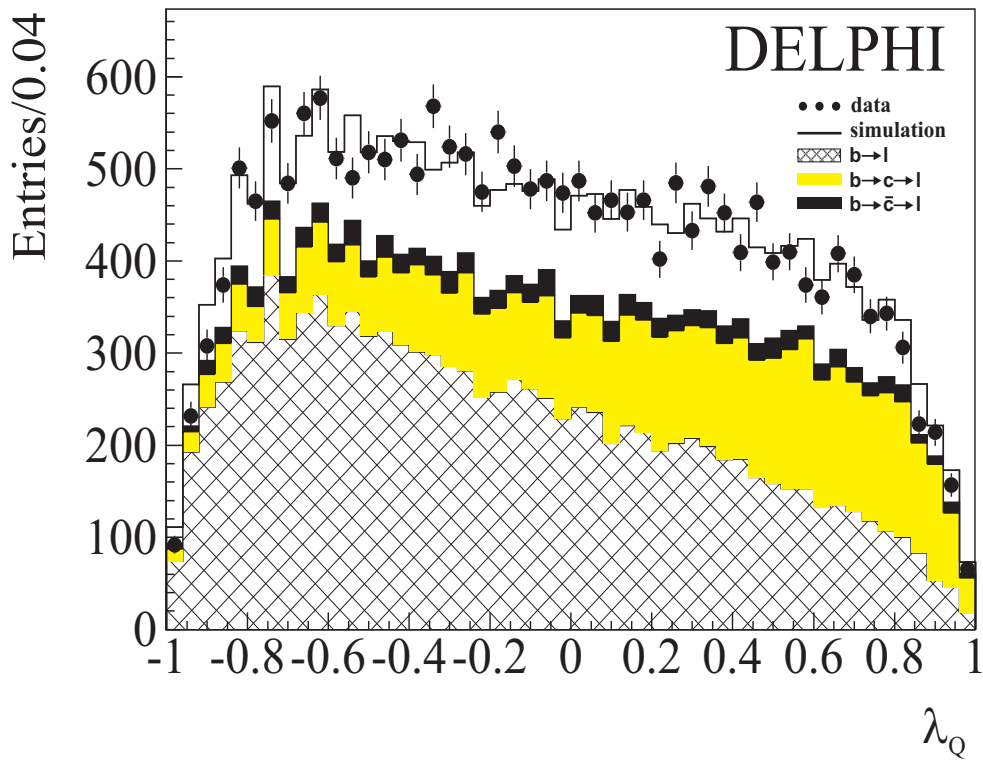,height=10cm,width=13cm}
\caption{Distribution of the charge correlation variable
\lm\ = $Q_\ell \cdot Q_{b}$
for data and simulation.
%The contributions from \btol\ events is given by the dark grey area, the
%\bctol\ (\bcbtol) contribution is represented by the light (darker) grey area.
}
\label{fig:jetch}
\ec 
\end{figure}

The charge of the lepton, $Q_\ell$, was compared to
 the one of the $b$ jet measured in the opposite hemisphere, $Q_b$. 
Neglecting mixing, the product \lm\ = $Q_\ell \cdot Q_b$
should  be, in case of perfect measurement,  -1/3  (+1/3) for leptons 
from direct (cascade) decays.
The charge of the $b$ quark was determined in each 
hemisphere by properly combining several quantities (jet charge, vertex charge,
 charge of any kaon or lepton from
$b$ decay, charge of leading fragmentation particles: a detailed description of
the method can be found in \cite{bosc}), such that \lm\ actually ranged between
 -1 (mostly \btol) and +1 (mostly \bctol). Figure 
\ref{fig:jetch} shows the 
\lm\ distribution for the data and simulation.  
The fraction of wrong charge assignment, for a given \lm\ range, depends on 
several quantities related both to the $b$ 
hadron production and decay mechanisms ($B$ mixing, fragmentation, lepton and 
$K$ production in $b$ decays, $b$ charged multiplicity,
 etc.) and to the detector performance (tracking, vertexing, particle 
identification), 
which are in some cases not well known. To reduce the systematic uncertainty, 
the fraction of correct tags 
% ($\eta_\lm $) 
was determined in the data, as explained in Section \ref{II_BR}.

For the previous analysis the charge correlation was only available  
for the di-lepton sample whereas \lm\ can be determined for all events: 
it should be noted however that the discrimination power of this variable is
smaller. Therefore the two analyses are complementary. 
Only 1994 data were used for this analysis.

\subsection{Determination of the branching fractions}
\label{II_BR}

The $b$ semileptonic branching fractions were obtained by means of a binned 
\XX  fit.
Leptons in the data and in the simulation were collected in two-dimensional 
bins, according to their 
\ks\ and \lm\ values, so as to exploit fully the discriminating power of
 the two variables. The \ks\ bins had adjustable widths, defined such as 
to correspond to  at least 40  entries in each bin. The range of the \lm\
values was divided into an even number ($N_\lm $) of bins of the equal width,
4 \lm and 25 $k^*$ bins were used.

\begin{figure} \bc
\epsfig{file=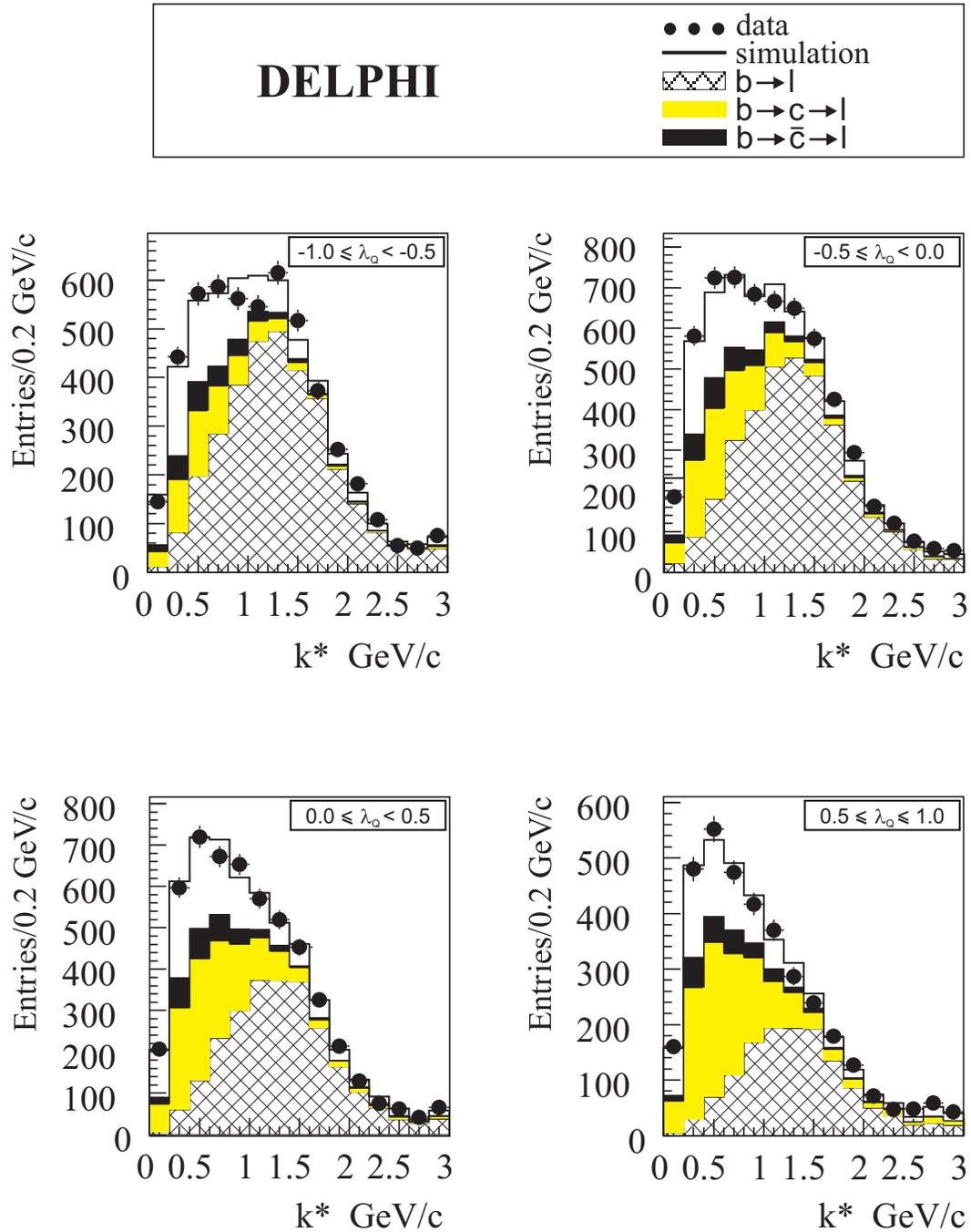,height=18cm,width=14cm}
\caption{Distribution of the lepton momentum in the b rest frame, \ks, shown
in different \lm\ bins.}
%The lower histograms represent the contributions from \btol\ (dark grey) 
%\bctol\ (light grey) and \bcbtol\ (darker grey).
%The four plots are obtained by requiring, respectively:
%$-1 < \lm < -0.5, -0.5 <\lm < 0., 0.< \lm < 0.5$ and  $0.5 < \lm < 1.$}
\label{fig:pspectra}
\ec 
\end{figure} 

Events in the simulation were assigned to one of the seven
classes described in Section \ref{sec:I_single} depending on their origin. 
Leptons from classes (d) to (g) were normalized to the data according to 
the number of hadronic events, known branching 
ratios and efficiency correction factors. 
The normalization factors for the classes (a), (b) and (c) were instead 
determined from the fit and used to compute the 
branching fractions for the direct (\btol) and cascade (\bctol, \bcbtol) 
semileptonic decays.
%In order to determine in addition 
%the BR(\bcbtol), the fit was then repeated using the normalization of 
%class (c)  as a further free parameter. 
Figure \ref{fig:pspectra} shows the fitted \ks\ distribution in four different 
\lm\ bins.

The fraction of correct charge tags in each \lm\ bin was determined while
 performing the fit. 
For this purpose, the total number of simulated events belonging to the 
class  $\alpha$ ($\alpha$=a,b,c) and falling in the $i^{th}$ ($j^{th}$)
\ks\ (\lm) bin (${\cal N}_{MC}^\alpha(i,j)$) were multiplied by a linear 
correction factor:
\ba
\nonumber {\cal N}^\alpha(i,j) ~=~ {{\cal N}_{MC}}^\alpha(i,j) ~\cdot~
 (1+\delta^\alpha_j)
\ea
where ${\cal N}^\alpha(i,j)$ is the number of data events in the same bin.
The $\delta$ coefficients would be zero if the simulation described the data
 perfectly. They were left as free parameters in the fit with the following 
constraints:
\bi
\item for a given \lm\ bin, $\delta$ does not depend on \ks\;
\item $\delta^{\rm a}_{j} = \delta^{\rm c}_{j} = \delta^{\rm b}_{k}$, 
where $k$ is the 
\lm\ bin with  opposite charge with respect to $j$ ($k$ = $N_\lm+1-j$);
\item 
$ \sum_{i,j} {\cal N}^\alpha(i,j) ~=~ \sum_{i,j} {\cal N}_{MC}^\alpha(i,j)$ 
for every  $\alpha$ 
\ei
The first requirement follows from the fact that the \lm\ value is computed 
in the hemisphere opposite to the lepton, and is therefore uncorrelated with 
the value of \ks\ and with all other lepton decay properties. 
The second constraint expresses the fact that leptons from direct and cascade
decays populate mainly cells that are symmetric with respect to \lm\ .
The third constraint ensures that the total number of events is conserved. 
Values of $\delta$ of about -7\% and  +4\% have been obtained for 
classes (a) and (b,c), respectively.
The fit results did not change significantly if the same correction was 
applied to the simulated leptons of the other classes (d-g). 

The procedure was performed separately for muons and electrons: consistent results were found. The \XX 
per degree of freedom was 0.95 for muons and 1.23 for electrons, 
There was no appreciable difference in the \XX 
when using different models to describe the lepton spectra.

\begin{table}[htb]
\begin{center}
\begin{tabular}{|l|ccc|}
\hline
          &  $\Brbl$ & $\Brbcl$ &$\Brbcbl$ \\
\hline
 $\Brbl$     & 1.00  &  0.017   & -0.228   \\
 $\Brbcl$    &       &  1.00    & -0.928   \\
 $\Brbcbl$   &       &          &  1.00    \\
\hline
\end{tabular}
\caption{Correlation matrix of statistical uncertainties in Analysis II.
}
\label{tab:II_corr}
\end{center}
\end{table}

\begin{table}
\begin{center}
\begin{sideways}%
\begin{minipage}{\textheight}
\begin{center}
\begin{tabular}{|l|c|c|c|c|}
\hline
Error Source &Range & $\Delta\Brbl$ &$\Delta\Brbcl$&$\Delta\Brbcbl$ \\
             &      & $10^{-2}$     & $10^{-2}$    & $10^{-2}$      \\ \hline

electron efficiency & $\pm  3. \%$  & $\mp$0.15  & $\mp$0.12 & $\mp$0.09 \\ 
misidentified electrons & & & & \\
and converted photons & 
             $\pm 8.\%, \pm 10\%$   & $\pm$0.01  & $\mp$0.03 & $\mp$0.08 \\
$\mu$ efficiency    & $\pm 2.5\%$   & $\mp$0.17  & $\mp$0.09 & $\mp$0.07 \\
misidentified $\mu$ & $\pm 6.5\%$   & $<$0.01    & $<$  0.01 & $\mp$0.07 \\ 
\hline
 $\varepsilon_c$      & $\pm 9 \%$  & $\pm$0.14  & $\pm$0.10 & $\pm$0.03 \\ 
 $\varepsilon_{uds}$  & $\pm 22\%$  & $\pm$0.03  & $\pm$0.02 & <0.01     \\ 
$\ell$-btag correlation&$\pm 1.\%$  & $\mp$0.05  & $\mp$0.11 & $\mp$0.03 \\
 $R_b$        & $0.21643\pm 0.00073$~\cite{ref:LEP99}
                       & $<$0.01    & $<$0.01   & $<$0.01   \\ 
 $R_c$        & $0.1694\pm 0.0038$~\cite{ref:LEP99}
                       & $\pm$0.01  & $\pm$0.01 & $\mp$0.01 \\ 
 binning      & $\pm$ 2 bins        & $\pm$ 0.05 & $\pm$0.05 & $\pm$0.05 \\    
\hline
 total experimental &               & $\pm$0.28  & $\pm$0.22 & $\pm$0.16 \\
\hline
$x_E(b)$      & $0.702 \pm 0.008$~\cite{ref:LEPHF}    & <  0.01    &  <  0.01  &  <  0.01 \\
$x_E(c)$      & $0.484 \pm 0.008$~\cite{ref:LEPHF}    & $\mp$0.02  & $\pm$0.02 &   <  0.01   \\
${b\rightarrow W\rightarrow D}\over{b\rightarrow W\rightarrow D_s}$
 &$(1.28^{+1.52}_{-0.61})$~\cite{ref:LEPHF} 
          &$\pm$0.03   &$^{+0.20}_{-0.11}$  &$^{-0.23}_{+0.13}$\\
BR($\btaul$)   &$(0.459\pm0.071)$\%~\cite{ref:PDG}
               & $\mp$0.01 & $\mp$0.04 & $\mp$0.10 \\
BR($\bpsill$)  &$(0.07\pm 0.01)$\%~\cite{ref:PDG}  
               & $\mp$0.02 & $\pm$0.01 & $\mp$0.02 \\
$\Brcl$        &$(9.85\pm0.32)$\%~\cite{ref:LEP99}  
               & $\mp$0.01 & <  0.01   & $\mp$0.02 \\ 
$\glcc$        &$(3.19\pm0.46)$\%~\cite{ref:LEP99}  
               & $<$0.01   & $<$0.01   & $<$0.01 \\
$\glbb$        &$(0.251\pm0.063)$\%~\cite{ref:LEP99}  
               & $<$0.01   & $<$0.01   & $<$0.01  \\
\hline
 total systematics &                & $\pm$0.28 & $\pm$0.28 & $\pm$0.27 \\
\hline
Semilept.mod.$\bl$\cite{ref:LEPHF}& ACCMM ($\mathrm{^{+ISGW}_{-ISGW**}}$)
                                    &$^{-0.33}_{+0.53}$ & $^{-0.27}_{+0.44}$
                                                        & $^{+0.56}_{-0.84}$\\
Semilept.mod.$\cl$\cite{ref:LEPHF}& ACCMM1($\mathrm{^{+ACCMM2}_{-ACCMM3}}$)  
                                    &$^{-0.08}_{+0.06}$ & $^{-0.22}_{+0.09}$
                                                        & $^{+0.07}_{-0.05}$\\
\hline
 total models &      &\large{$^{-0.34}_{+0.53}$} & \large{$^{-0.35}_{+0.50}$} 
                                                 & \large{$^{+0.56}_{-0.84}$}\\
\hline
\end{tabular}
\end{center}
\caption[bla]{Summary of systematic uncertainties in the 
analysis of lepton vs jet charge.
Ranges given  in \% correspond to  relative variations around the
central value.}
\label{tab:II_sys}
\end{minipage}
\end{sideways}
\end{center}
\end{table}

The final results, averaged between electrons and muons, are:
%\ba
%\nonumber BR(\btol) &=& (10.87 \pm 0.13 (stat) \pm 0.29 (syst) _{+0.46}
%^{-0.31} (model) ) \% \\ 
%\nonumber BR(\bctol) &=& (~8.71 \pm 0.25 (stat) \pm 0.49 (syst) ^{+0.42}
%_{-0.64} (model) )\% 
%\ea
%for the fit with two parameters and 
\ba
\nonumber BR(\btol) &=& (10.78 \pm 0.14 (stat) \pm 0.28 (syst) _{+0.53}
^{-0.34} (model) ) \% \\ 
\nonumber BR(\bctol) &=& (~7.59 \pm 0.69 (stat) \pm 0.28 (syst) _{+0.50}
^{-0.35} (model) ) \% \\
\nonumber BR(\bcbtol) &=& (~2.00 \pm 0.49 (stat) \pm 0.27 (syst) ^{+0.56}
_{-0.84} (model)  \% 
\ea
%for the other one. 
The average correlation matrix for the statistical uncertainties 
is shown in Table \ref{tab:II_corr}.
The breakdown of the systematic uncertainties  for the fit is 
presented in Table~\ref{tab:II_sys}. The variation of the \ks\ resolution 
causes small differences in the  bins population which are included in the 
binning error.

%**************************************************************************

\section{Analysis III: Measurement of semileptonic $b$ decays by
     applying a multitag method} 
\label{sec:anal_III}
\par 
A measurement of BR($b \to \mu$) and BR($b \to c (\bar{c}) \to \mu$) 
using data collected with the DELPHI detector between 1992 and 1995 is 
presented here. 
Muons were  identified as described in Section  \ref{sec:muid}.

In this analysis the contributions of $uds$, $c$ and $b$ flavours were 
separated in an inclusive way using  a multitag method which 
used almost all the hadronic events, because it was based 
on a flavour deconvolution without the need for any further cuts.  
One important by-product of the method was a systematic and independent 
analysis of the muon background; as this study cannot be simply applied 
at electrons due to the presence of photon conversions, 
all the analysis has been performed with muons only.

The selection of the hadronic events was the same as in 
Section~\ref{sec:eventsel} except that 
 five charged particles instead of seven  were required to select the event, 
and the event thrust axis was required to satisfy 
      $|\cos \theta_{th}| < 0.75$.

The total numbers of selected events both in real and simulated data
are shown in Table~\ref{Table:III_numev}.

\subsection{Flavour tagging}
\label{sec:III_btag}
The $uds$, $c$ and $b$ events were separated using the
multivariate analysis which was previously applied to the 
$\Gamma_{b\bar{b}}/ \Gamma_{had}$ determination \cite{ref:btag}. 
In each event hemisphere defined with respect to the thrust axis, 
a set of discriminating variables, called discriminators, were calculated,
using lifetime information and event shape variables. These were combined in  
the {\it multivariate flavour tagging} algorithm~\cite{ref:NIM} and
the {\it flavour confidence} algorithm~\cite{ref:btag}.
The outputs of these two algorithms were then combined as in~\cite{ref:btag}.
By applying cuts to the combined discriminator and, as in~\cite{ref:btag}, 
using the enhanced impact parameter tag to define the b-tight category,
each hemisphere was classified  
in one of the following six categories: uds-loose, uds-tight, charm, b-loose, 
b-standard and  b-tight, numbered from 1 to 6 respectively. 

The 6 hemisphere categories provide 21 corresponding event categories 
and hence 21 equations from which the 18-3 independent probabilities,
$\varepsilon^j_i$, of classifying a hemisphere of flavour $j$
in category $i$ ($j = b,c,uds$ and $i=1,...,6$) and the 3-1 independent $R_j$
values, the fractions of flavour $j$ hemispheres in the whole sample,
might be determined from a fit to the data. But in practice, because of a 
rotational ambiguity in the system, 3 additional inputs have to be given. 
As in \cite{ref:btag}, these were chosen to be $R_c$ and the probabilities
$\varepsilon^{uds}_{b-tight}$ and $\varepsilon^{c}_{b-tight}$ of classifying 
charm and uds hemispheres in the b-tight category.

In this analysis the main output of this step is the determination of the 
probabilities $\varepsilon^j_i$, and hence the flavour content of the different
hemisphere categories, rather than that of $R_b$.
The cuts on the combined discriminators
have therefore been re-optimized with respect to \cite{ref:btag}.
The cut on the extended impact parameter tag, however, was kept unchanged 
in order to keep the values of $\varepsilon^{uds}_{b-tight}$ and
$\varepsilon^{c}_{b-tight}$ unchanged from those determined in~\cite{ref:btag}.
The value of $R_b$ obtained was $R_b=0.21741\pm0.00065$ (stat).

The two main features of this method are 
the minimal correlation between hemispheres 
(because the event vertex was computed independently in each hemisphere) 
and 
the direct measurement of the tagging efficiencies and of the 
flavour composition from the data.
Since 1994, due to the introduction of double sided 
silicon detectors measuring $z$ as well as $r\phi$, 
a better $b$-flavour tagging has been achieved.

\begin{center}
\begin{table} \bc
\begin{tabular}{|c|c|c|c|c||c|}
\cline{2-6}
\multicolumn{1}{c|}{ }     & 1992    & 1993    & 1994    & 1995    & Total   \\
\hline
Simulated                  & $1 \thinspace 369 \thinspace 156$ & $1 \thinspace
232 \thinspace 678$ & $2 \thinspace 275 \thinspace 552$ &  $712 \thinspace 868$
 & $5 \thinspace 590 \thinspace 254$ \\
Real data                  &  $486 \thinspace 357$ &  $471 \thinspace 437$ & 
 $971 \thinspace 448$ &  $467 \thinspace 809$ & $2 \thinspace 397 \thinspace 051$\\
\hline
\end{tabular}
\caption{Total numbers of selected events for Analysis III}
\label{Table:III_numev}
\ec
\end{table} 
\end{center}

\par
 
\subsection{Flavour deconvolution}

The aim of the flavour deconvolution was to extract the spectra 
of the muon variables $p$, $p_t^{in}$ and $p_t^{out}$ for each flavour,
where $p$ is the momentum of the  muon candidate 
and $p_t^{in}$ and $p_t^{out}$ are its transverse momentum  with respect
to the jet axis including ($p_t^{in}$) or excluding ($p_t^{out}$) the muon 
in the definition of the jet. 
Hereafter any of these variables will be referred to as $z$. 
The inputs to the flavour deconvolution were the distributions of these 
variables for each of the six categories defined in the previous section:
the category assigned to an identified muon was the category found by the 
tagging in the opposite hemisphere, in order to  avoid 
correlations between the hemisphere tagging and the presence of the muon.
\par 
A $\chi^2$ was then constructed using the number $n_{i}^{\mu}(z)$ of identified 
muons in a given category, $i$, in an interval of $z$:  
\begin{eqnarray}
  \label{eqn:decon-2}
  \chi^2 = \sum_i \frac{ \left ( n_i^\mu(z) - 
                        N_{hem} \left (\sum_j \varepsilon_i^j R_j D^{\mu}_j(z) 
                        \right ) \right )^2}
                       { n_i^{\mu}(z)} 
\end{eqnarray}
where $N_{hem}$ is the total number of hemispheres, $R_j$ and $\varepsilon^j_i$ 
are the flavour fractions and tagging probabilities extracted from the data as 
just explained above, and $D^\mu_j(z)$ is the
spectrum of the $z$ variable for flavour $j$ extracted from the flavour
deconvolution. 
The above formula neglects correlations between the hemisphere tagging and muon
selection efficiencies in opposite hemispheres.  
\par
The minimization of this $\chi^2$ function leads to a set of three linear
equations for each $z$ bin, where the three unknowns are the components
of the spectrum in each flavour: $D^{\mu}_{uds}(z)$, $D^{\mu}_{c}(z)$, 
$D^{\mu}_{b}(z)$. These quantities, and their errors, were computed by solving 
these equations.

Thus, as a result of the deconvolution, the spectra of identified muons in the 
different flavours were obtained.
They can be written as a function of the different sources of muons: 
  \begin{eqnarray}
   \label{eqn:decon-3}
%     \begin{split}
     n^{\mu}_{uds}(z) = & N_{hem} R_{uds} D^{\mu}_{uds}(z) & = 
                           n_{uds}^{bg\mu}(z) \nonumber \\ 
     n^{\mu}_{c}(z)   = & N_{hem} R_c D^{\mu}_c(z) & = 
                           n_c^{p\mu}(z) + n_c^{bg\mu}(z) \\
     n^{\mu}_{b}(z)   = & N_{hem} R_b D^{\mu}_b(z) & = 
                           n_b^{p\mu}(z)                + n_b^{bg\mu}(z) \nonumber 
%                          n_b^{p\mu}(z) + n_b^{c\mu}(z)+ n_b^{b\mu}(z) \nonumber 
%     \end{split}
  \end{eqnarray}

\noindent
where $n_{uds}^{bg\mu}(z)$, $n_c^{bg\mu}(z)$ and $n_b^{bg\mu}(z)$ are the
distributions of \textit{background} muons for different flavours, 
%$n_b^{c\mu}(z)$ is the distribution of muons coming from \textit{cascade} 
%$b\to c\to\mu$ and $b\to\bar c\to\mu$ decays of the $b$, 
and $n_{c}^{p\mu}(z)$ and $n_{b}^{p\mu}(z)$ are the distributions of 
{\em prompt} muons coming from $c$ and $b$ decays respectively. 
%In the $b$ flavour,
%the total number of $prompt$ muons $n_b^{g\mu}(z)$ can be expressed
% as $n_b^{g\mu}(z) = n_b^{p\mu}(z) + n_b^{c\mu}(z)$. 
\par
This method of flavour deconvolution can also be applied to other kinds 
of particles and observables. For example, the deconvolution can 
be applied to all charged particles.  The distributions obtained
with charged particles are interesting results in themselves,
but are here used only to compute the backgrounds $n_c^{bg\mu}(z)$ and
$n_b^{bg\mu}(z)$ from $n_{uds}^{bg\mu}(z)$, 
as described in the next section.  

\subsection{Background extraction and hadron misidentification probability}
\label{sec:III_back}
\par

In this analysis, a background muon was defined as any particle identified as 
a muon that either was not a muon, or was a muon but
from a light hadron (mainly pion or kaon) decay. Following this definition, 
all identified muons in $uds$ events were taken as background.
%the fraction of muons from the decay of a heavy quark produced from gluon 
%splitting was assumed to be negligible. 
The misidentification probability, $\eta_{uds}$, was then defined 
as the fraction of charged particles identified as muons in $uds$ events: 
\begin{equation}
  \label{eqn:decon-4a}
  \eta_{uds}(z) = \frac{n_{uds}^{\mu}(z)}{n_{uds}^{tk}(z)}
\end{equation}
where $n_{uds}^{tk}(z)$ is the spectrum of charged particles 
with the same kinematic cuts as the muons in the $uds$ sector. 

This can be expressed as:
\begin{equation}
\label{eqn:eta}
\eta_{uds}(z) = \eta^{\pi}(z)  {f}^{\pi}_{uds}(z) + 
              \eta^{K}(z)    {f}^{K}_{uds}(z) +
              \eta^{\mu}(z)    {f}^{\mu}_{uds}(z) +
              \eta^{o}(z)    {f}^{o}_{uds}(z)
\end{equation}
where $\eta^{\pi}(z)$ and $\eta^{K}(z)$ are the misidentification 
probabilities for pions and kaons, 
${f}^{\pi}_{uds}(z)$ and ${f}^{K}_{uds}(z)$ 
are the fractions of pions and kaons for the $uds$ flavour, 
${f}^{\mu}_{uds}(z)$ is the fraction of muons 
coming from $\pi$ and $K$ decays in flight 
and $\eta^{\mu}(z)$ is their identification efficiency, 
and  ${f}^{o}_{uds}(z)$
and $\eta^{o}(z)$ are respectively the fraction and the misidentification 
probability of
\textit{other} charged particles, which are mainly protons.
 The fractions for
the different flavours and particles have been measured in DELPHI
\cite{ref:schyns}, and agree with the predictions obtained with
the JETSET simulation program and used in this analysis.
The specific misidentification probabilities 
($\eta^{\pi}(z)$, $\eta^{K}(z)$, ...) were supposed to be flavour 
independent but, 
since the fractions of these particles are not the same in $uds$, $c$ 
and $b$ events, a  different misidentification probability was 
evaluated for each flavour ($\eta_{uds}$, $\eta_c$ and $\eta_b$).
Equation (\ref{eqn:eta}) was used to extract  $\eta^{\pi}(z)$, taking 
$\eta_{uds}(z)$ from the data and $\alpha_{K\pi}=\eta^K(z)/\eta^\pi(z)$,
$\eta^\mu(z)$ and $\eta^o(z)$ from the simulation. 
Then, from equations analogous to (\ref{eqn:eta}) written for $c$ and $b$ 
flavours, $\eta_c$ and $\eta_b$ were calculated.
\par
The misidentification probabilities obtained with this method were 
compared with those obtained using a tight anti-$b$ cut in 
Section~\ref{sec:muid}, and good agreement was observed. 
%(see reference \cite{ref:dnote} for more details). 

Once the misidentification probability for each flavour was computed,
the numbers of background muons per hemisphere for a variable $z$,
i.e. the $n^{bg\mu}(z)$ in (\ref{eqn:decon-3}), were obtained
by multiplying them by the number of charged particles per hemisphere for 
each flavour.
Subtracting these contaminations from the muon candidates per hemisphere, 
it was possible to determine the distributions of prompt muons. 

\subsection{Fitting of prompt muon distribution}
\label{sec:III_fit}

\par
In order to measure the branching fractions $BR(b \to \mu)$
and $BR(b \to c(\bar{c}) \to \mu)$, the following $\chi^2$ function was then 
minimized: 
\begin{equation}
  \label{eqn:decon-20}
     \chi^2 = \sum_{i=1}^m 
           \frac{ \left ( n_b^{p\mu}(z_i)-n_b^{p\mu,th}(z_i) \right )^2}
                        {  n_b^{p\mu}(z_i) }  
\end{equation}
where $m$ is the number of bins, $n_b^{p\mu}(z_i)$ is 
the distribution of prompt muons measured as described above,
and  $n_b^{p\mu,th}(z_i)$ is a model expectation which can be written as:
\begin{eqnarray}
  \label{eqn:decon-21} 
   n_b^{p\mu,th}(z) & = & N_{hem} R_{b}
     \left( 1 + BR(g \to b\bar{b}) \right) \nonumber \\
     &\times& \left[ \epsilon_{b \to \mu}(z) P_{b \to \mu}(z) 
                                               BR(b \to \mu)+ \right. 
      \left. \epsilon_{b \rightarrow c (\bar{c}) \to \mu}(z) 
                              P_{b \to c(\bar{c}) \to \mu}(z)  
                              BR(b \to c(\bar{c}) \to \mu) \right]~~~ \\
    & ~  & +n^{\mu}_{b \to  \tau \to \mu}(z)+ 
        n^{\mu}_{b \to  J/\psi \to \mu}(z)+
        n^{\mu}_{g \to c\bar{c} \to \mu}(z) \nonumber
\end{eqnarray}
where $BR(b \to \mu)$ and $BR(b \to c(\bar{c}) \to \mu)$ are the only unknowns,
and $P_{b \to \mu}(z)$ and $P_{b \to c(\bar{c}) \to \mu}(z)$ are the
true spectra of muons coming from $b\to\mu$ and $b\to~c(\bar{c})\to\mu$
decays which were taken from different models:
for the central value, the ACCMM model has been used for $b \to \mu$ decays 
and the ACCMM1 model for $c \to \mu$ decays. 
The additional terms $n^{\mu}_{b \to \tau \to \mu}(z)$, 
$n^{\mu}_{b \to  J/\psi \to \mu}(z)$ and 
$n^{\mu}_{g \to c\bar{c} \to \mu}(z)$ are the contributions to
\textit{prompt} muons coming from $b \to  \tau \to \mu$, 
$b \to  J/\psi \to \mu$ and $g \to c\bar{c} \to \mu$ decays, respectively. 
The shapes of these distributions have been taken directly from the simulation,
 but the recommendations of~\cite{ref:LEPHF} have been followed 
for their normalizations.

The factors $\epsilon_{b \to \mu}$ and $\epsilon_{b \rightarrow c (\bar{c}) 
\to \mu}$ are global efficiency factors which contain the product of
the efficiencies  for the momentum cut ($p>3$ GeV$/c$)
and the muon geometrical acceptance, evaluated for each of the two considered 
channels, and the muon identification efficiency.

\begin{figure}
 \begin{center}
   \epsfig{file=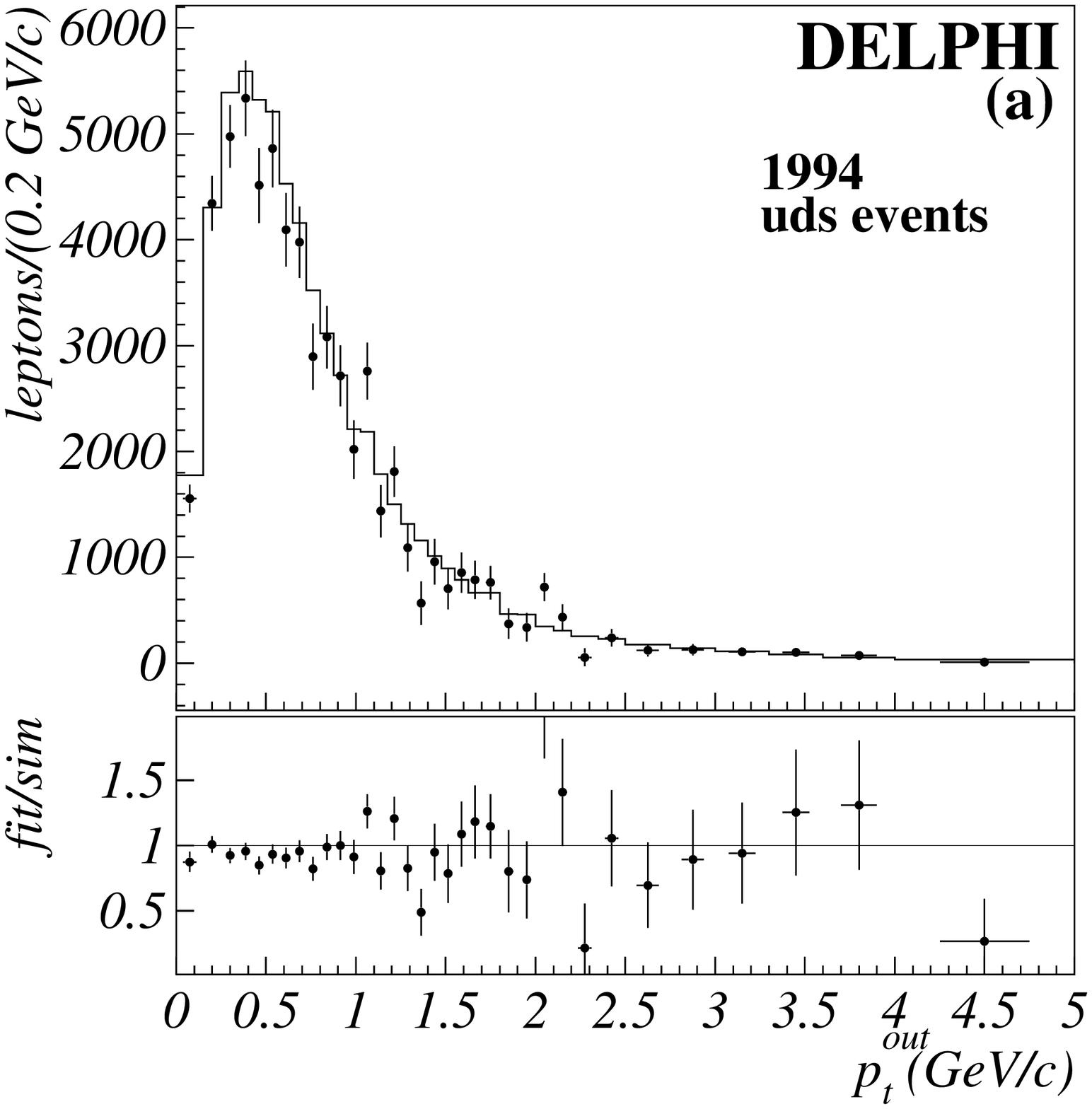,height=7.cm,width=7.cm} 
   \epsfig{file=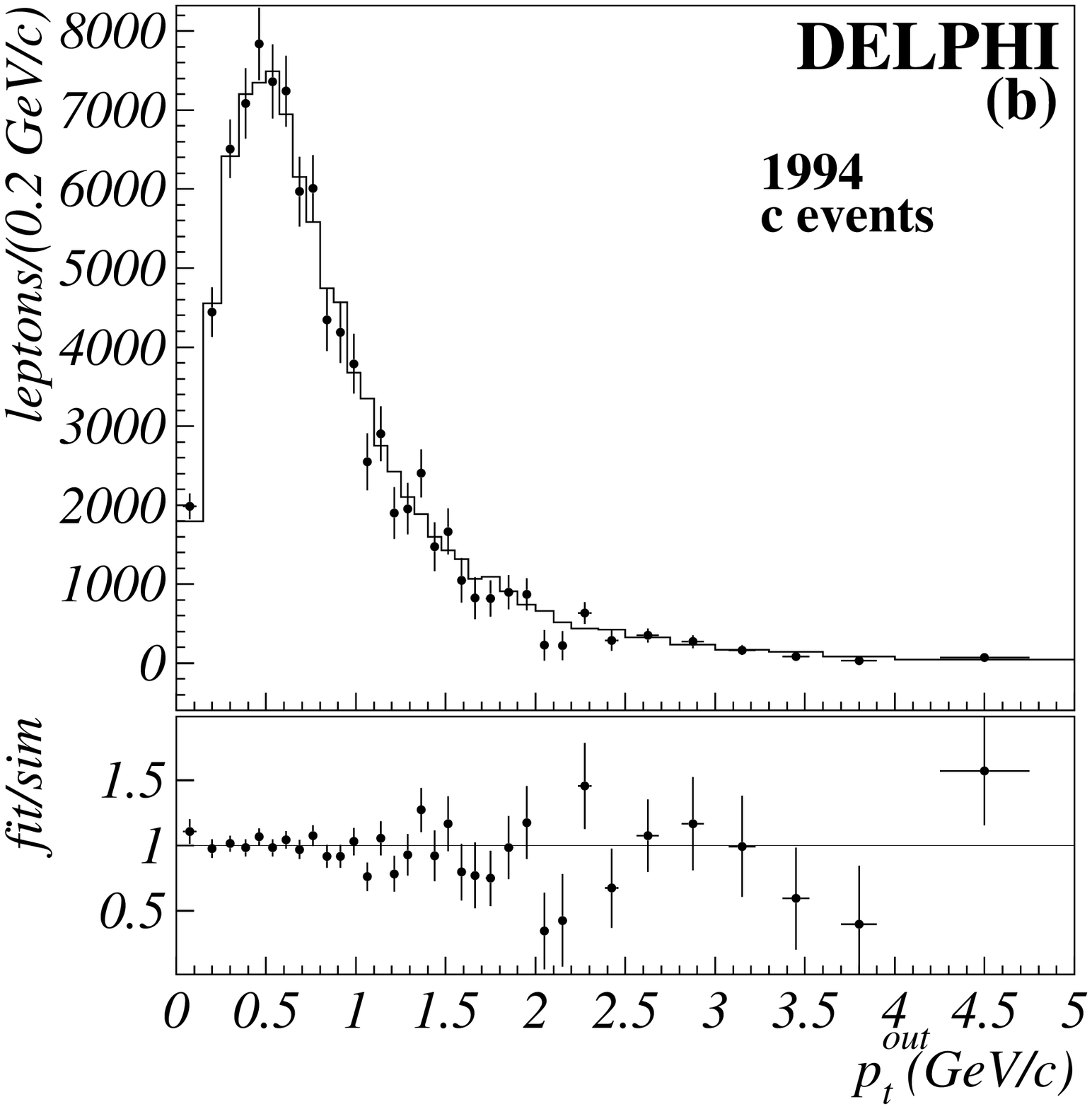,height=7.cm,width=7.cm} \\
   \epsfig{file=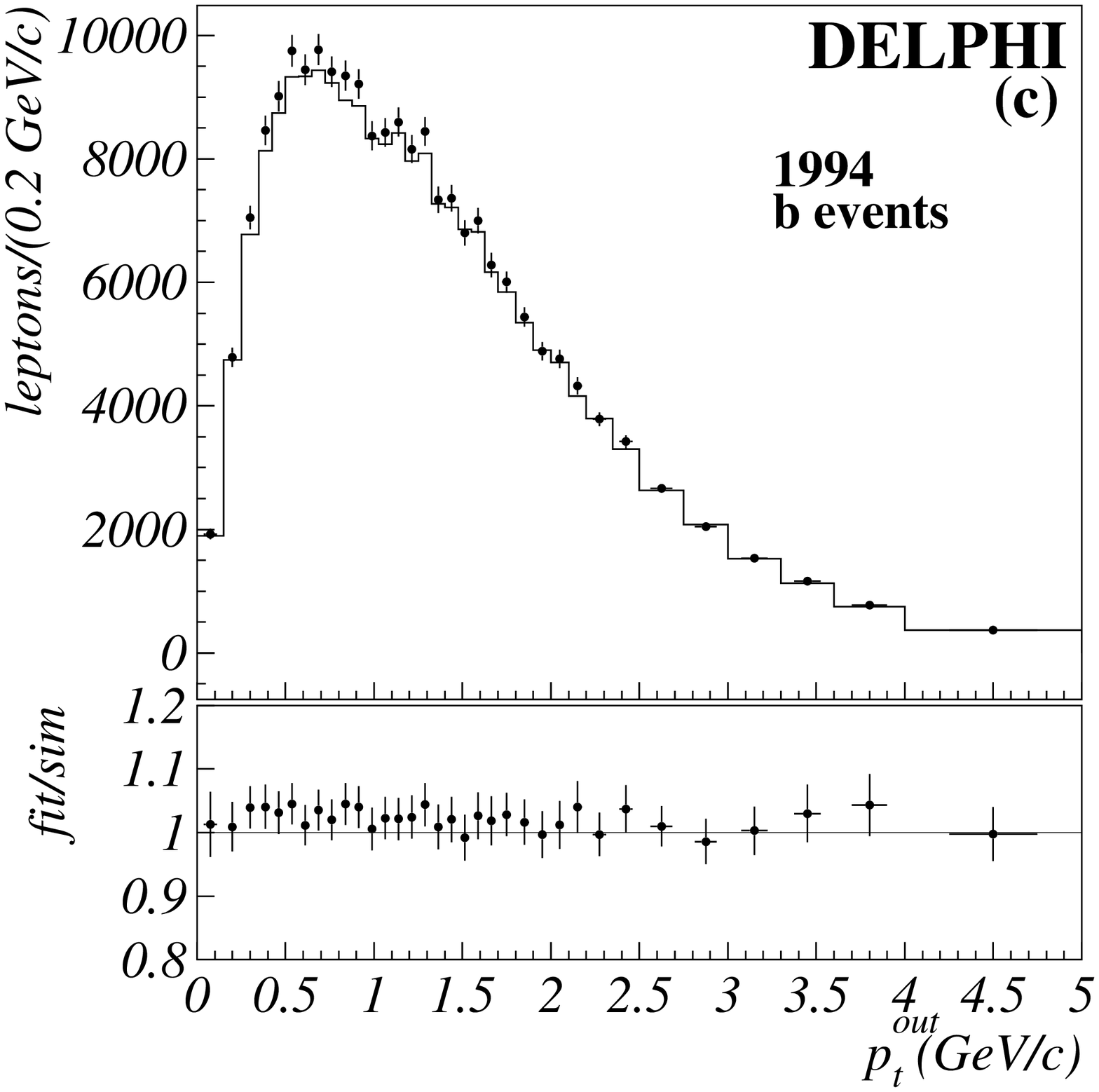,height=7.cm,width=7.cm} 
   \caption{Separation of $p_t^{out}$ spectra of candidate muons between the
            three flavors. The upper part of each plot compares 
            the results of the deconvolution in simulated data (points) 
            with the generated spectra (solid line);
            the lower part shows the ratio between these two
            distributions.}
    \label{fig:decon-2}
 \end{center} 
\end{figure} 

\subsection{Results and systematic errors}
\label{sec:III_results}

The semileptonic branching fractions were obtained minimizing the binned 
$\chi^2$ of equation (\ref{eqn:decon-20}).
In order to check the validity of the method, a test was performed using
simulated  data. Figure \ref{fig:decon-2} shows a comparison between the
muon $p_t^{out}$ distributions at generation level and after deconvolution.
A small discrepancy is visible in the $b$ sample. 
The difference between the generated values 
of the semileptonic branching fractions and the fit results were found to 
be 0.8\% and 1.4\% for the direct and cascade muons, respectively.  
These differences take into account the approximations used in the analysis. 
They were used to correct the results obtained with data
and were also taken as systematic error contributions.

\begin{table}[ht]
  \begin{center}
  \begin{tabular}{|c|c|c|c|c|}
   \cline{3-5}
    \multicolumn{2}{c|}{} & $b\to\mu$ & $b\to c(\bar{c})\to\mu$ &$\chi^2/dof$\\
    \multicolumn{2}{c|}{} &    (\%)   &    (\%)                &             \\
    \hline
%    \multirow{3}{14mm}{1992} 
            & $p$         & 10.78 $\pm$ 0.28 & 9.22 $\pm$ 0.46 & 25.38/27 \\
     1992   & $p_t^{in}$  & 10.79 $\pm$ 0.25 & 9.68 $\pm$ 0.42 & 25.20/32 \\
            & $p_t^{out}$ & 10.75 $\pm$ 0.22 & 9.81 $\pm$ 0.37 & 22.75/32 \\ 
    \hline
%    \multirow{3}{14mm}{1993} 
            & $p$         & 10.77 $\pm$ 0.29 & 9.24 $\pm$ 0.50 & 30.62/27 \\
     1993   & $p_t^{in}$  & 10.68 $\pm$ 0.25 & 9.77 $\pm$ 0.45 & 30.02/32 \\
            & $p_t^{out}$ & 10.63 $\pm$ 0.22 & 9.78 $\pm$ 0.40 & 41.62/32 \\
    \hline
%    \multirow{3}{14mm}{1994} 
            & $p$         & 10.77 $\pm$ 0.18 & 9.60 $\pm$ 0.25 & 43.05/27 \\
     1994   & $p_t^{in}$  & 10.73 $\pm$ 0.16 & 9.43 $\pm$ 0.28 & 27.74/32 \\
            & $p_t^{out}$ & 10.62 $\pm$ 0.14 & 9.54 $\pm$ 0.24 & 37.16/32 \\
    \hline
%    \multirow{3}{14mm}{1995} 
            & $p$         & 10.76 $\pm$ 0.29 & 9.69 $\pm$ 0.45 & 18.82/27 \\
     1995   & $p_t^{in}$  & 10.72 $\pm$ 0.24 & 9.86 $\pm$ 0.41 & 24.21/32 \\
            & $p_t^{out}$ & 10.67 $\pm$ 0.21 & 9.93 $\pm$ 0.36 & 39.26/32 \\
    \hline
    \end{tabular}
    \caption{Fit result for the  real data (the errors are only statistical).}
    \label{tab:bslrd}
  \end{center}
\end{table}

The results obtained applying the fitting procedure to the real data are 
shown in Table~\ref{tab:bslrd}.
%The errors are only statistical and vary, for each year, for the different 
%fitting variables. The reason of this behaviour can be
%found in the distribution of the different sources of \textit{prompt} muons. 
It can be seen that some variables, 
which separate the different contributions in different regions,
are more discriminant than others.
For the transverse momentum, $b \to c(\bar{c}) \to \mu$ events are
concentrated at low values, while $b \to \mu$ events are mainly situated at
high transverse momentum. On the other hand in the $p$ distribution,  
in the low momentum region both contributions are of similar importance. 
Thus the errors on the semileptonic branching fractions 
extracted using the transverse momentum distributions are expected to be 
lower than those obtained using the momentum distribution.

\begin{figure}
 \begin{center}
  \epsfig{file=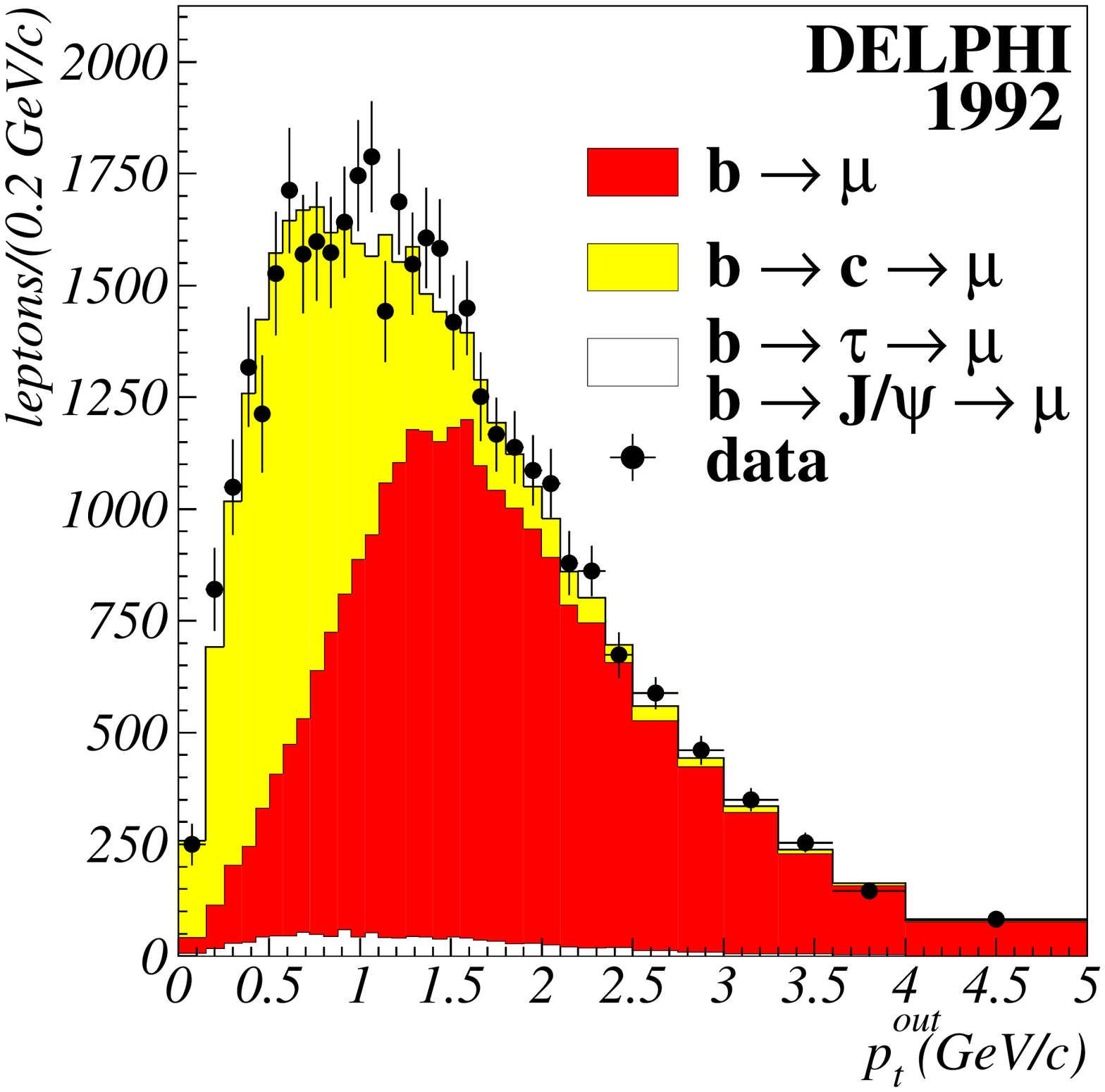,height=0.4\textwidth, width=0.4\textwidth}
  \epsfig{file=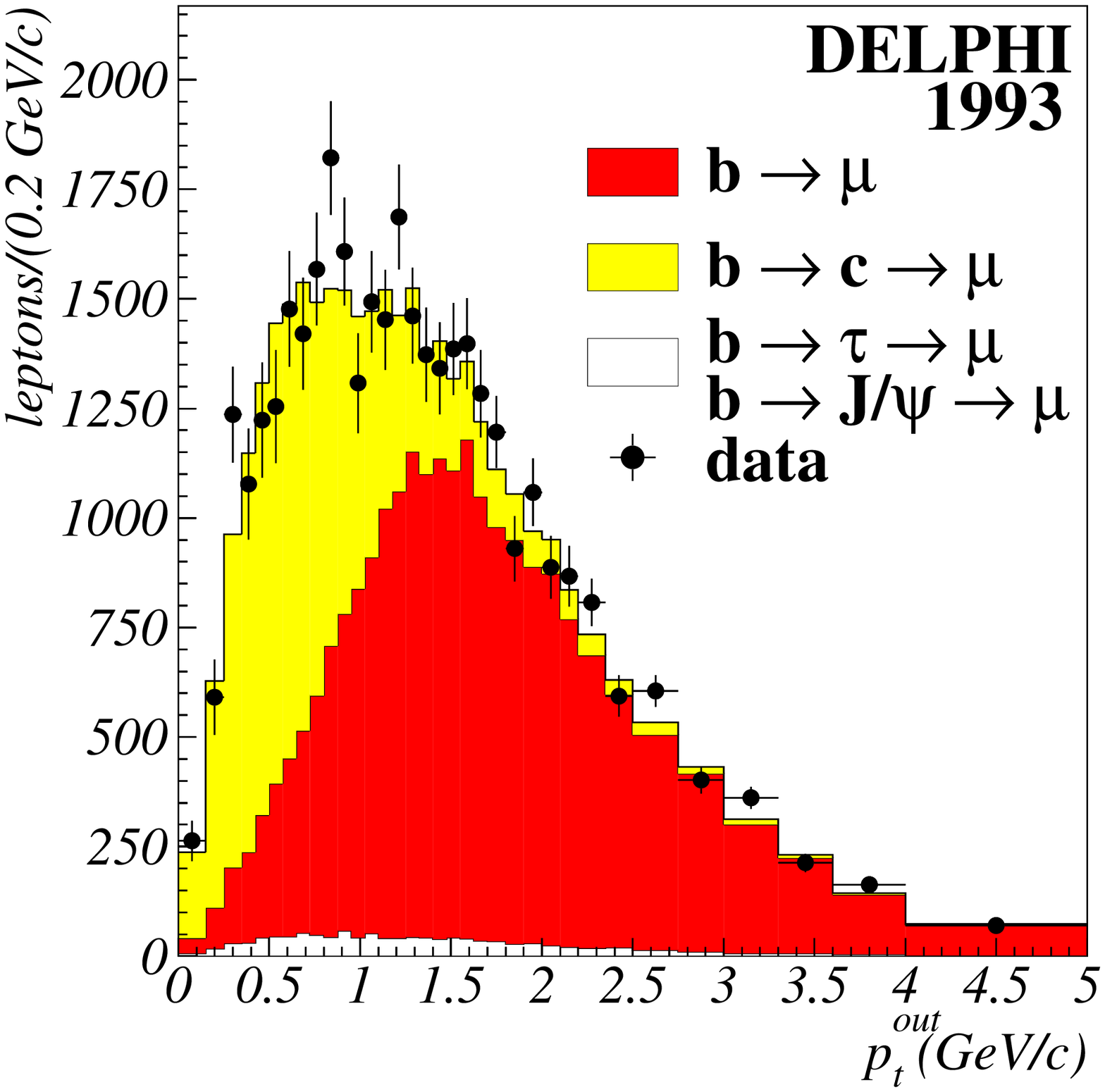,height=0.4\textwidth, width=0.4\textwidth} \\
  \epsfig{file=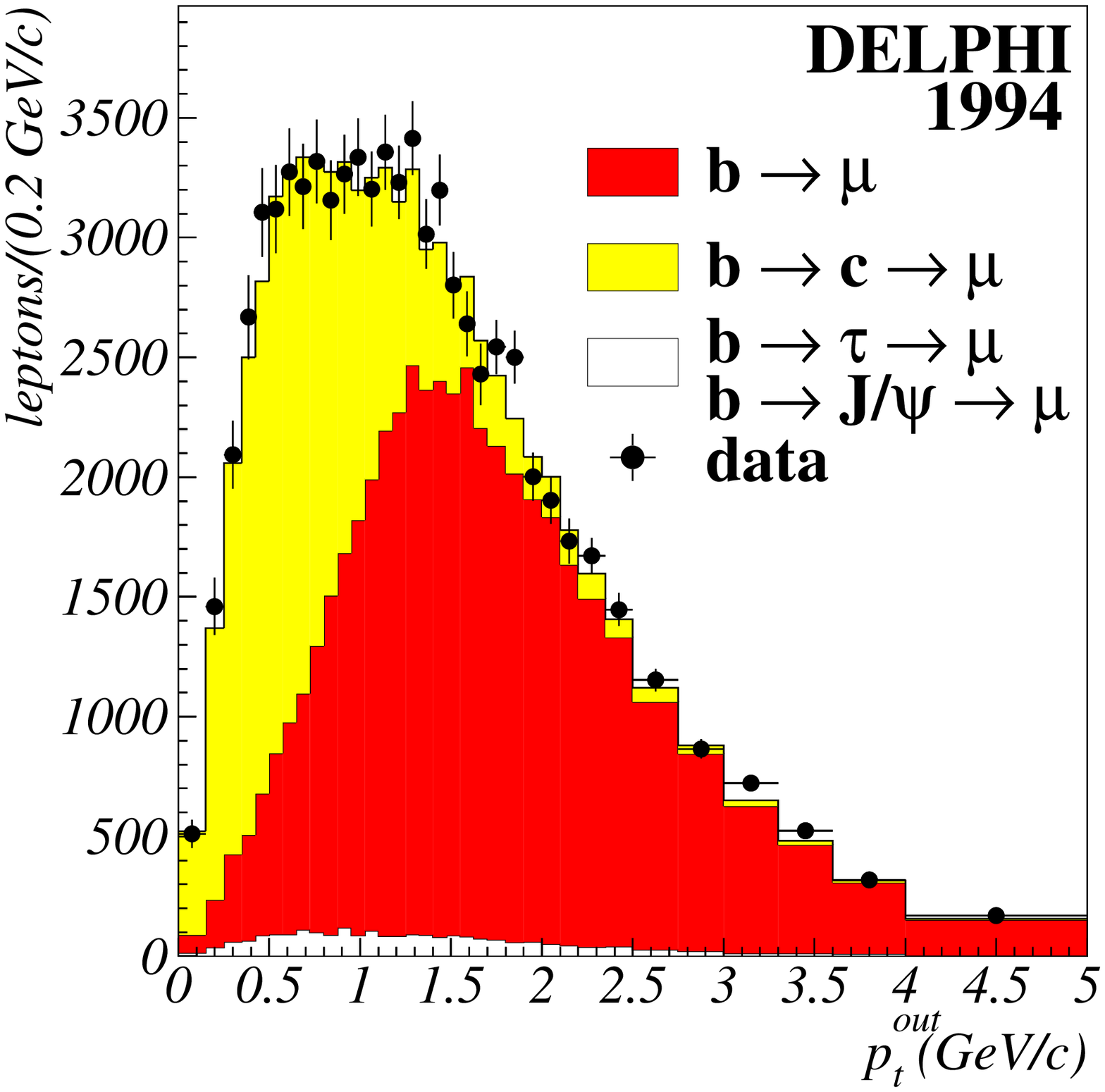,height=0.4\textwidth, width=0.4\textwidth}
  \epsfig{file=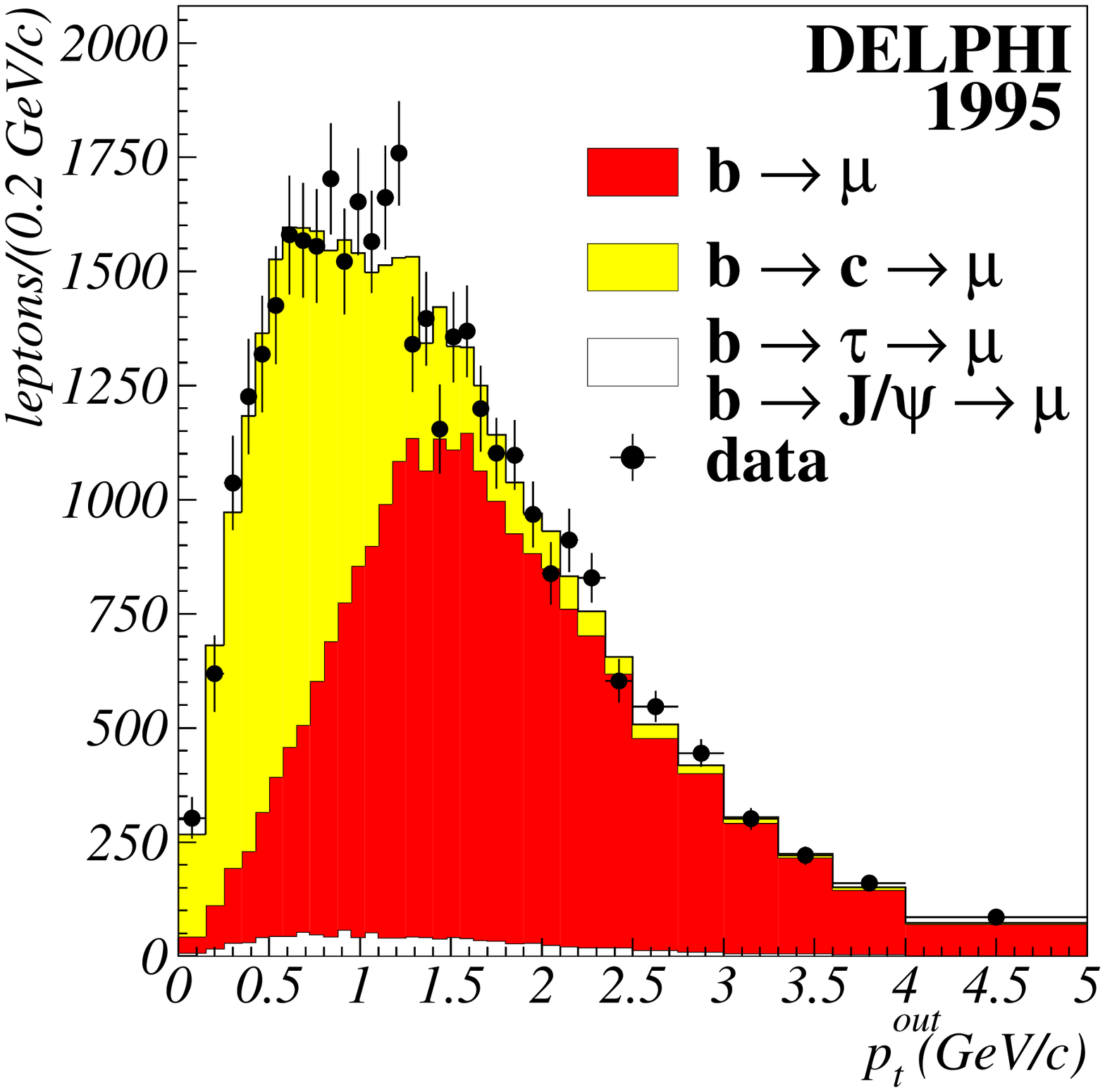,height=0.4\textwidth, width=0.4\textwidth} 
  \caption{Comparison of the $p_t^{out}$ distributions of prompt muons for 
the $b$ flavour in real data (dots) with the distributions obtained using 
the semileptonic branching fractions (histograms).
The contributions of different processes are displayed.}  
    \label{fig:bslrd-3}
  \end{center}  
\end{figure}

Once the $b$ semileptonic branching fractions have been fitted, it is possible
to calculate the $b \to \mu$ and the $b \to c(\bar{c}) \to \mu$  spectra
using the model  spectra $P_{b \to \mu}(z)$ and 
$P_{b \to c(\bar{c}) \to \mu}(z)$. The results are displayed  in  
Figure \ref{fig:bslrd-3} for each year of data taking. 
The small contributions coming from the 
$b \to  \tau \to \mu$ and  $b \to  J/\psi \to \mu$ decay channels,
taken directly from the simulation, are also shown.

\renewcommand{\arraystretch}{1.2}
\begin{table}[h]
 \begin{center}
  \begin{tabular}{|c|c|c|}
   \hline       
     Source & $\Delta(b \to \mu)$ & $\Delta(b \to c(\bar{c}) \to \mu$) \\
   \hline
   \hline
    muon efficiency ($\pm 2.5\%$)    & $\mp$ 0.190 & $\mp$ 0.182 \\
   \hline
    ${f}^{\pi}_b$ ($\pm 2\sigma$)& $\mp$ 0.004 & $\mp$ 0.008 \\
    ${f}^K_b$   ($\pm 2\sigma$)  & $\mp$ 0.002 & $\mp$ 0.007 \\
    ${f}^{\mu}_b$ ($\pm 2\sigma$)& $\pm$ 0.003 & $\pm$ 0.009 \\
    ${f}^o_b$   ($\pm 2\sigma$)  & $\mp$ 0.001 & $\mp$ 0.001 \\
   \hline  
    $\eta^{\pi}$  ($\pm 2\sigma$)  & $\mp$ 0.022 & $\mp$ 0.120 \\
    $\alpha_{K\pi}$ ($\pm 2\sigma$)& $\pm$ 0.008 & $\mp$ 0.035\\
    $\eta^{\mu}$  ($\pm 2\sigma$)  & $\mp$ 0.004 & $\mp$ 0.004\\
    $\eta^{o}$    ($\pm 2\sigma$)  & $\mp$ 0.001 & $\mp$ 0.001\\
   \hline 
    $R_b = 0.2170\pm0.0009$        &  $<0.01$  & $<0.01$ \\
    $R_c = 0.1734\pm0.0048$        &  $<0.01$  & $<0.01$    \\
    $\varepsilon^{uds}_{b-tight}$ ($\pm 15\%$) 
                                   & $\pm$ 0.023  & $\pm$ 0.010\\
    $\varepsilon^c_{b-tight}$ ($\pm 7\%$) 
                                   & $\pm$ 0.007  & $\pm$ 0.028 \\
   \hline    
    Variable                       & $\pm$ 0.080 & $\pm$ 0.150\\
    Muon quality                   & $\pm$ 0.082 & $\pm$ 0.082\\
    Binning                        & $\pm$ 0.078 & $\pm$ 0.079\\
    Bias of the method             & $\pm$ 0.080 & $\pm$ 0.136\\
   \hline 
    MC statistics                  & $\pm$ 0.088 & $\pm$ 0.163\\
   \hline
    $x_E(b) =  0.702 \pm 0.008$
                                   & $\pm$ 0.093  & $\pm$ 0.165 \\
   \hline 
%    $BR(b \to u)$ ($2.6 \pm 0.2$ )\%
%                                   & $\mp$ 0.008 & $\mp$ 0.014\\
    $BR(c \to \ell) = (9.85\pm0.32)$\%~\cite{ref:LEP99}  
                                   & $\mp$ 0.001 & $\mp$ 0.002\\
    $BR(b \to \tau \to \ell^{-}) = (0.459\pm0.071)$\%~\cite{ref:PDG}  
                                   & $\mp$ 0.014 & $\mp$ 0.096\\
    $BR(b \to J/\psi \to \ell^{-} \ell^{+})=(0.07\pm 0.01)\%$~\cite{ref:PDG}  
                                   & $\mp$ 0.018 & $\mp$ 0.011\\
    $BR(g \to c\bar{c}) = (3.19\pm0.46)$\%~\cite{ref:LEP99}  
                                   & $\pm$ 0.009 & $\pm$ 0.010\\ 
    $BR(g \to b\bar{b}) = (0.251\pm0.063)$\%~\cite{ref:LEP99}  
                                   & $\mp$ 0.033 & $\mp$ 0.043\\
   \hline 
    total systematic               & $\pm$ 0.28 & $\pm$ 0.41\\
   \hline 
    $b \to \ell$  ~~~$ACCMM^{+ISGW}_{-ISGW**}$    
                                   & $^{-0.35}_{+0.43}$ 
                                   & $^{+0.52}_{-0.48}$ \\ 
    $c \to \ell$  $ACCMM1^{+ACCMM2}_{-ACCMM3}$ 
                                   & $^{-0.11}_{+0.11}$ 
                                   & $^{-0.12}_{+0.02}$ \\
   \hline 
    total models              & $_{-0.37}^{+0.44}$ & $^{+0.52}_{-0.49}$\\
   \hline 
  \end{tabular}
 \end{center}
 \caption{Analysis III: Systematic uncertainties (\%) for $BR(b \to \mu)$ and 
$BR(b \to c(\bar{c}) \to \mu)$} 
 \label{tab:III_sys} 
\end{table}
\renewcommand{\arraystretch}{1.0}

%%%Sources of systematic uncertainties have been grouped into seven different 
%%%categories. The first four categories are related to detector effects: 
%%%muon identification and
%%%misidentification, hemisphere tagging, analysis method and Monte Carlo 
%%%statistics. The other three take into account 
%%%the theoretical knowledge on decays: fragmentation parameters, assumed 
%%%branching ratios and decay models.
%%%In table \ref{tab:III_sys} all these errors
%%%for $b \to \mu$ and $b \to c(\bar{c}) \to \mu$ as well as the range of
%%%variation are shown.
%%%%%%%%%%%%%%%%%%%%%

%%%%***********  New  version  ********** Comentada 24/Julio/2000*****

Sources of systematic uncertainties have been grouped into several different 
categories. 
%The first four categories are related to detector effects: 
%muon identification and misidentification, hemisphere tagging, analysis method 
%and Monte Carlo statistics. 
Here we comment briefly on the features that are specific 
to this analysis:

\begin{itemize}

%\item {\em muon id. efficiency:} First we consider the error due to the 
%inaccuracy of the muon identification efficiency: in this case an error of 
%2.5\% has been considered in the evaluation of the systematic error.

\item {\em muon misidentification:} The independent determination of the
background distributions in this analysis is affected by

\begin{itemize} 

\item the values of $f^{\pi}_{b}$, $f^{K}_{b}$, $f^{\mu}_b$ and $f^{o}_b$ 
which are the fractions of pions, kaons, muons (coming from $\pi$ and $K$  
decays in flight), and other charged particles in $b$ events; the central
values were taken from JETSET and the errors ($\sigma$) in the table are 
taken from \cite{ref:schyns}; $2\sigma$ ranges are taken to conservatively
cover the degree to which the DELPHI data~\cite{ref:schyns} corroborated the
JETSET values.

\item the misidentification probabilities specific to the particles such as
$\eta^{\pi}$, which has been evaluated from $\eta_{uds}$, the ratio 
$\alpha_{K\pi}$, which has been taken from simulation, and 
$\eta^{\mu}$ and  $\eta^{o}$, whose contribution is small and has 
also been taken from simulation.
\end{itemize}
\item {\em hemisphere tagging:} in order to use the multivariate method,
three parameters had to be fixed externally: $R_c$ and the probabilities 
$\varepsilon^{uds}_{b-tight}$ and $\varepsilon^{c}_{b-tight}$; the
variations of the latter probabilities correspond to their systematic
uncertainties as evaluated in~\cite{ref:btag}. 
%{\it add remarks on - correlations?? - $R_b$ value??}
The variation corresponding to the difference between the $R_b$ value 
resulting from this analysis and the reference value used from the other three 
analyses was found to be negligible.  

 \item {\em analysis method:} here the effects of different choices 
made in our analysis are considered, namely 
(i) the choice of the variable (i.e. $p$, $p_t^{in}$ or $p_t^{out}$), 
(ii) the effect of using a looser muon selection,
(iii) the influence of changing the number of bins of our variables, and
(iv) the effect of the bias shown in Figure \ref{fig:decon-2} and discussed 
above.

\end{itemize}

For each year the results obtained with the three variables were
averaged assuming complete correlation in the statistical error.
After averaging over the four years, taking into account the
correlations between the systematic errors,  the results are:
\vskip 0.25cm
\begin{eqnarray*}
               BR(b \to \mu) & = & (10.71 \pm 0.11 (stat) \pm 0.28 (syst)
                                          ^{-0.37}_{+0.44} (model)) \% \\
BR(b \to c(\bar{c}) \to \mu) & = &  (~9.62 \pm 0.19 (stat) \pm 0.41 (syst) 
                                          ^{+0.52}_{-0.49} (model)) \% 
\end{eqnarray*}

%**************************************************************************

\section{Analysis IV: Measurement of semileptonic $b$ decays from
inclusive $b$-hadron reconstruction and charge correlation}
\label{sec:anal_IV}

In this analysis the charge correlation between the $b$ quark and the lepton 
produced in its decay was used  to measure the  semileptonic decay rates of  
$b$-hadrons.
The two different cases leading to the like charges, 
direct decay (\btol) 
and ``upper decay vertex''  ( \bcbtol), 
were separated on the basis of different lepton momentum regions.

To use the charge correlation method,  
$b$-hadrons containing a $b$-quark, $H_b$, needed to be separated from those 
containing a $\bar{b}$-quark, $H_{\bar{b}}$. 
This separation was accomplished in four steps: 1) by isolating $b\bar b$ 
events, 2) by reconstructing the $b$-hadron  decay vertex, 3) by 
identifying the tracks from the $b$-hadron  vertex and finally 4) 
by estimating the hadron charge.  The details of these four steps are
described below in section \ref{sec:AIV_method}.1 to \ref{sec:AIV_method}.4.
% and can be found in references \cite{BSAURUS,FYJERU}.
 After the separation, the sign of the charge
of the $b$-quark and that of the lepton were compared, and
each lepton was classified into ``like-sign'' or ``opposite-sign'' categories.
The fit of the like-sign spectrum was performed assuming the sample was 
composed of \btol and \bcbtol decays, whereas the opposite-sign
spectrum assumed only \bctol decays.

\subsection{$B$ reconstruction and separation between $H_b$ and $H_{\bar{b}}$}
\label{sec:AIV_method}
\subsubsection{Event selection}

Hadronic events were selected in the same manner as described in
Section \ref{sec:eventsel} and  the event thrust axis was required
to be within the region $|\cos\theta_{thrust}| < 0.75$ to
ensure a good $b$-tagging efficiency.
In addition, good detector operating conditions were required for 
all detectors, including the RICH detector, used for hadron identification. 
Such requirements led to the selection of
$644 \thinspace 792$ and $223 \thinspace 082$ events in 1994 and 1995 data 
taking periods, respectively.
Each event was then divided into two hemispheres
with respect to the thrust axis, and the combined $b$-tagging
algorithm described in Section \ref{sec:btag} was applied to
select hemispheres enriched in $b$-hadron content.  The number of 
tagged hemispheres which contain a $b$ quark was estimated
using the same technique as in Section \ref{sec:I_single}. 
A slightly different cut on the combined $b$-tagging variable 
was used in this analysis, obtaining in simulation the following  
$c$ and $uds$ efficiencies:
$ \varepsilon_b = (42.50 \pm 0.06 (stat))\%  $,
$ \varepsilon_c =     (3.01  \pm 0.02 (stat))\%  $,
$ \varepsilon_{uds} = (0.329 \pm 0.003(stat))\% $.
This led to the purity of all $b$-tagged hemisphere being
$(92.6 \pm 0.3 (stat))\%$.

For each $b$-tagged hemisphere, lepton candidates were selected
in the opposite hemisphere using the same criteria as in 
Section~\ref{sec:leptid}.  
 This method avoids introducing a bias on the relative fraction of the 
 different $b$-hadron species in the hemispheres where lepton
 candidates were selected.

\subsubsection{Reconstruction of the $b$-hadron vertex}
\label{sec:anal_IV_recon}
In reconstructing the $b$-hadron decay vertex, the rapidity method  
presented in reference \cite{DELBSTAR} was used.
The reference axis for the rapidity calculation was defined by the jet
direction obtained using the LUCLUS algorithm with the transverse
momentum as the distance between jets and the parameter
$d_{join}$ set to 5 GeV/$c$.
The rapidity of each charged and neutral particle with respect to the 
reference axis was calculated, the particles outside the central 
rapidity window of $\pm 1.5$ were selected as $b-$hadron decay products
and used to reconstruct the secondary vertex.
A raw  $b$-hadron mass and energy were computed from the sum of the 
momentum vectors of the selected particles in the jet.
These values were corrected depending on the reconstructed mass 
and hemisphere energy.
This led to a relative energy resolution of about $7\%$ for $75\%$ of 
the $b$ hadrons
which constitute a Gaussian distribution, with the remainder making  a  tail 
at higher energies.

\subsubsection{Identifying tracks from the  $b$-hadron decay vertex}
\begin{figure}
\begin{center}
\epsfig{file=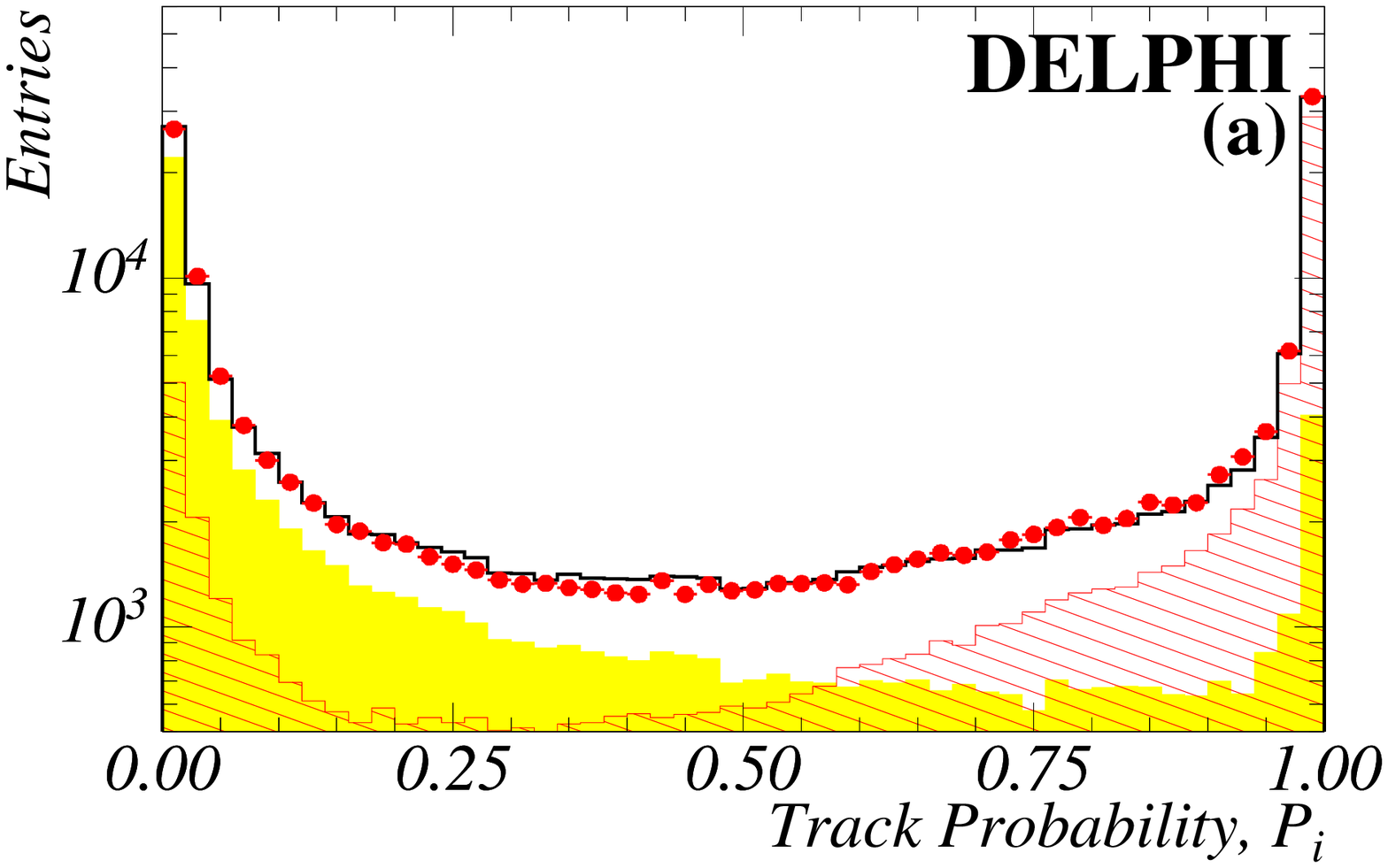,width=140mm,height=90mm}
\epsfig{file=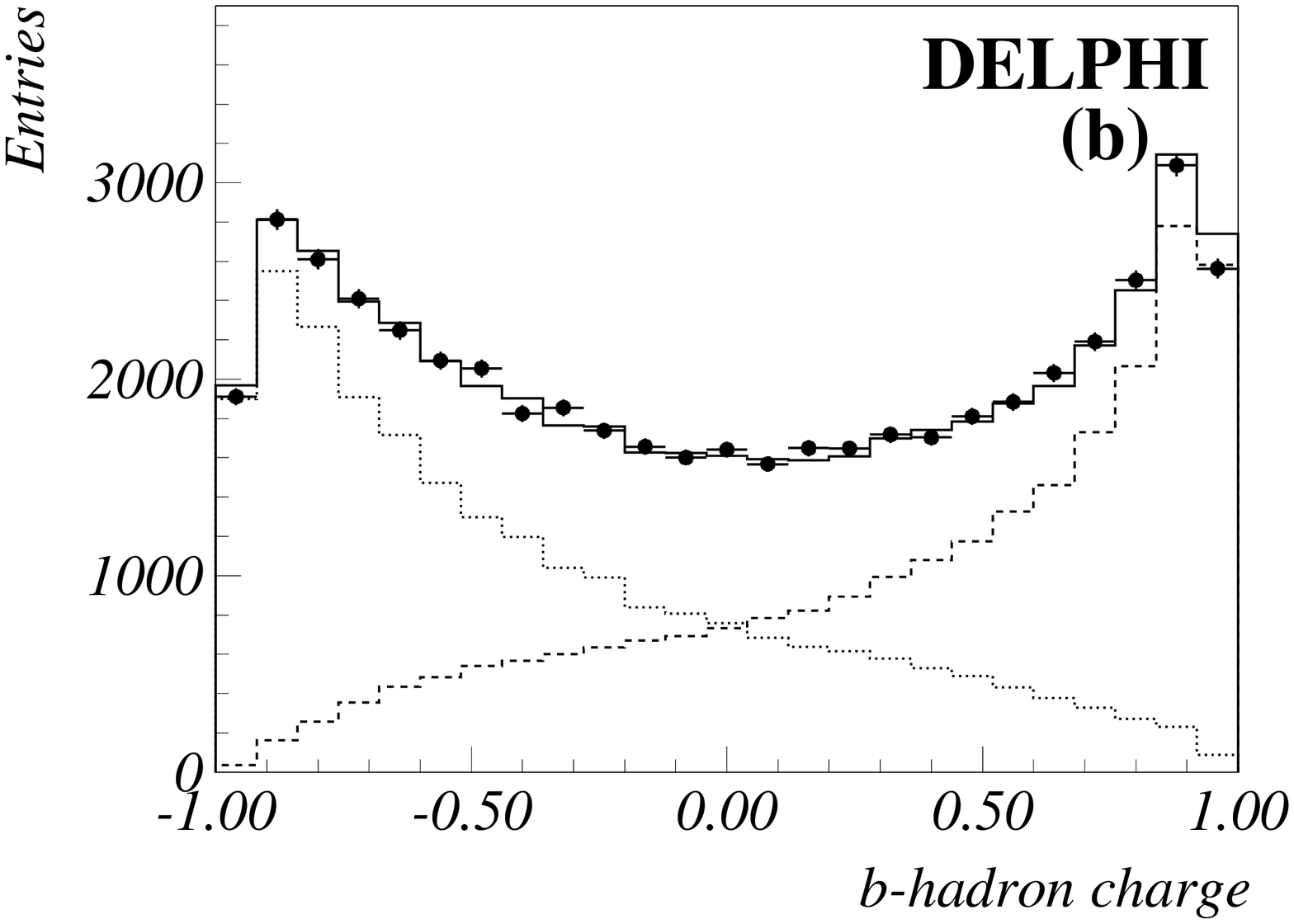,width=140mm,height=90mm}
\end{center}
\caption{{\bf (a)} Distribution of the track 
probability of real data compared
to the simulation shown in log scale: Solid (hatched) area represents
the tracks from fragmentation (from $b$-hadron decay).
{\bf (b)} 
$b$-hadron charge for real data compared
to the simulation: dotted (dashed) curve  represents
the $H_b$ ($H_{\bar{b}}$).}
\label{fig:fynet}
\end{figure}

For each charged particle a probability, $P_i$, that the particle originated
from a $b$-hadron decay rather than from fragmentation was calculated using a
 neural network.
It took into account the particle rapidity and 
momentum, its probability to originate from the primary vertex, its probability
to originate from the fitted secondary vertex, the flight distance and 
the energy of the hemisphere.
Figure \ref{fig:fynet}(a) shows 
the comparison between the real data and the  simulation.

\subsubsection{Classification of $H_b$ and $H_{\bar{b}}$}

For each hemisphere,  the vertex charge $Q_B = \sum Q_i P_i$ and
its uncertainty  
$\sigma_{Q_B}=\sqrt{\sum P_i(1-P_i)}$
were calculated by using the  probability, $P_i$, and the charge, 
$Q_i$, of each particle.
%%the charge $\sigma_{Q_B} = \sqrt{\sum P_i ( 1-P_i)}$
%%was also computed.  
These values, combined with the charge of the identified kaon from $b$-hadron
decay, the jet charge and the charge of the leading fragmentation
particle were fed into a neural network to classify a $b$-hadron
into $H_b$ or $H_{\bar{b}}$.  
The jet charge was defined as:
$ Q_{jet} = \frac{\sum Q_i\cdot|\overrightarrow
        {p_i}\cdot\overrightarrow{t}|^{\kappa}}
        {\sum |\overrightarrow{p_i}\cdot
        \overrightarrow{t}|^{\kappa}}$,
where $\overrightarrow{t}$ is the direction of the thrust axis and  
$\overrightarrow{p_i}$ is the momentum of the track.
Using simulation, the weighting exponent $\kappa$ was tuned to 
optimize the probability of correctly
assigning the charge of $b$-hadron and was chosen to be 0.6.
Figure \ref{fig:fynet}(b) shows the
comparison between the real data and the simulation.

\subsection{Measurements}
\subsubsection{Lepton selection}
The lepton identification was performed as in Section  \ref{sec:leptid}.
In addition, the lepton candidate was required to 
originate from the $b$-hadron  decay vertex by requiring its probability $P_i$ 
to be larger than 0.5.

For each selected lepton, its momentum $k^*$, in the  $b$-hadron rest frame, 
was calculated using the  $b$-hadron four-momentum calculated in 
Section \ref{sec:anal_IV_recon}.
%Like-sign and opposite-sign histogram were then filled with each lepton.
Since the average resolution on $k^*$ is 0.1 GeV$/c$, the $k^*$ distribution 
was chosen with a bin width of 0.2 GeV$/c$ to reduce migration
effects.

\subsubsection{Fitting and results}
The $k^*$ distributions of leptons classified as
``like-sign'' and ``opposite-sign''
were compared to the expected spectra from simulation and the branching 
fractions were extracted by means of a $\chi^2$ binned  fit.
The background contributions which may arise from non-$b$ events,
non-$b$-decay products and wrongly identified leptons were estimated from 
the simulation and  subtracted.
%The differential decay rate for an identified lepton, 
%${{dN}\over {dk^*}}(H_{\bar{b}}\to l^\pm X)$
%was then calculated as:
%\begin{equation}
%{{dN}\over{dk^*}}(H_{\bar{b}}\to l^\pm (k^*) X) =
%        (N^{data}_i - N^{MC,bg}_i) \cdot \frac
%       {N^{MC,gen}_i}{N^{MC,rec}_i} \frac{1}{k^*_{i+1}-k^*_i}
%       \cdot \frac{1}{N(H_{\bar{b}})}
%\end{equation}
%with bin definition $k^*_i < k^* < k^*_{i+1}$.
%%The background from the incorrectly determined 
%%charge of the $b$-quark was first estimated from the simulation
%%and used in the fit.  The results were then used to
%%adjust the background level and branching fractions.  
Any incorrectly determined charge of the $b$-quark led to 
the misclassification of leptons from like-sign to opposite-sign
and vice versa.  The amount of misclassified leptons was
first estimated from the simulation and used in the fit
of the lepton spectra.  
The fraction of each type of decay obtained from the fit was then used
to adjust the amount of misclassified leptons.  This process
was repeated until the fitting results converged.

%The Peterson fragmentation parameter \cite{ref:pet83},$\epsilon_b$,
%was determined from repeatedly adjusting $\epsilon_b$ and recalculating
%the corresponding lepton branching fractions until convergence was
%reached.  The mean $b$-hadron energy was found to be 
%$x_E = 0.704\pm0.003(stat)$.  

\begin{figure}
\begin{center}
\epsfig{file=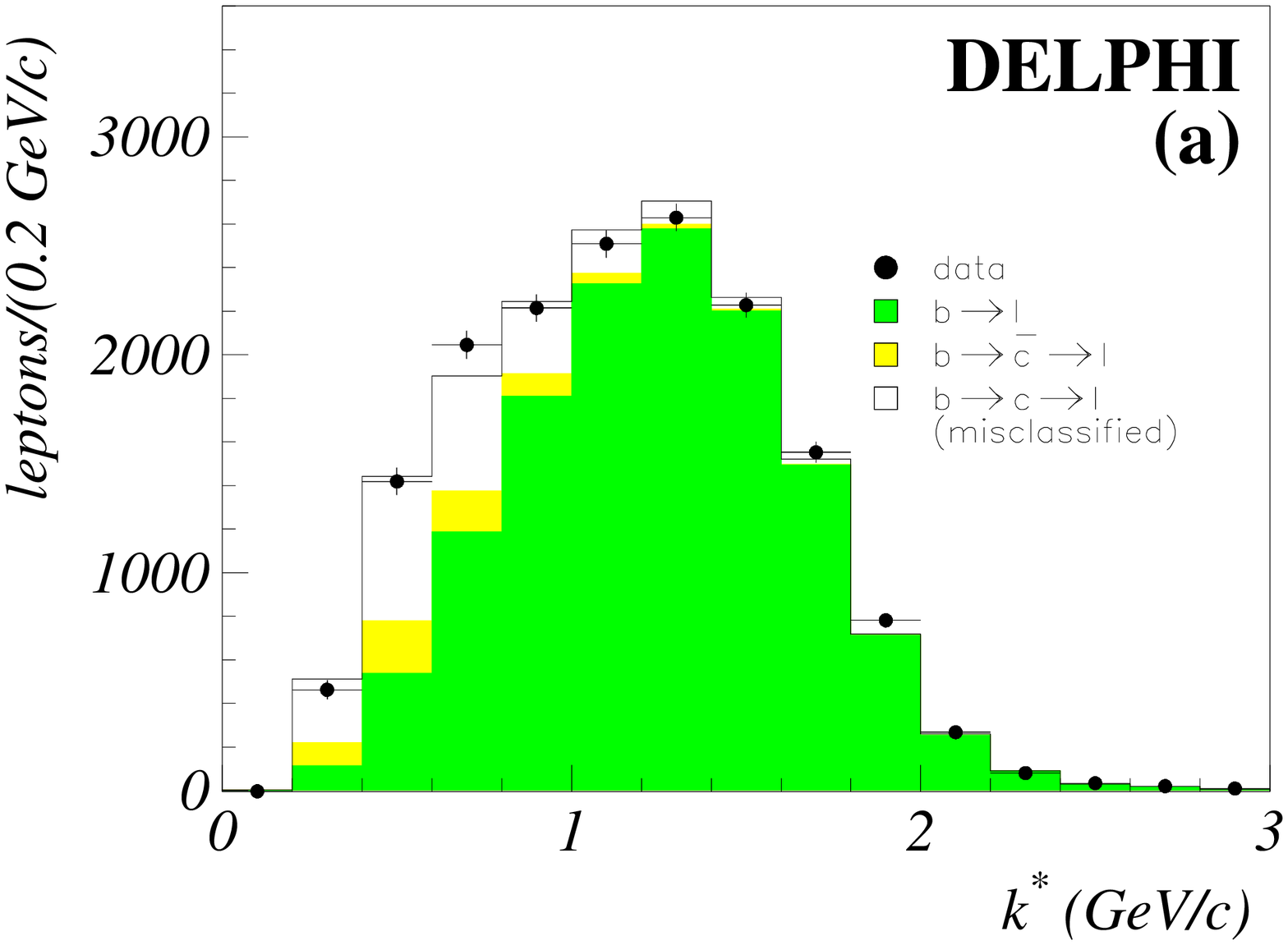,width=130mm}
\epsfig{file=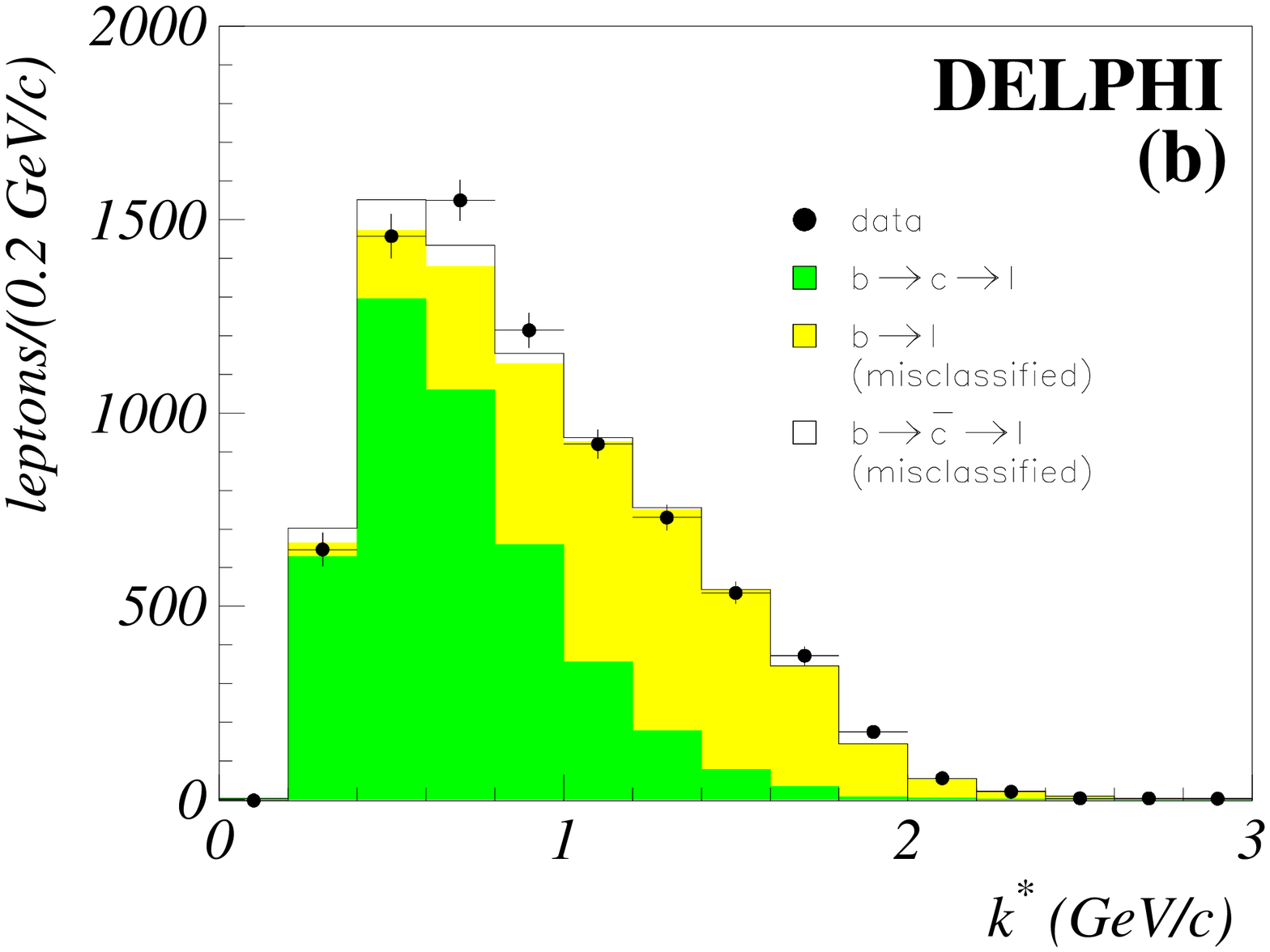,width=130mm}
\end{center}
\caption{Lepton momentum spectra in the $b$-hadron rest frame. 
Plot {\bf (a) ((b))} shows the result of the fit with the ACCMM
model to the like-sign (opposite-sign) sample .}
\label{fig:fyspectra}
\end{figure}

The following results have been obtained, and 
Figure \ref{fig:fyspectra} shows the results of the fit using the
 ACCMM model, where the uncertainties  are only statistical: 

\begin{center}
\begin{tabular}{|l|c|c|c|}
\hline
             &  1994           &   1995         & combined \\
\hline
BR$(b\to\ell^-)(\%) $ &       $10.78\pm0.18$&$10.67\pm0.30$&$10.75\pm0.15$  \\
BR$(b\to c\to\ell^+)(\%)$ &    $8.02\pm0.31 $&$7.92\pm0.52$ &$7.99\pm0.27$  \\
BR$(b \to\bar{c}\to\ell^-)(\%)$&$1.33\pm0.32 $&$1.36\pm0.50$ &$1.34\pm0.30$  \\

\hline
\end{tabular}
\end{center}

The following correlation matrix was found:
\begin{center}
\begin{tabular}{|l|ccc|}
\hline
          &  $\Brbl$ & $\Brbcl$ &$\Brbcbl$ \\
\hline
 $\Brbl$     & 1.00  & -0.077   & -0.350   \\
 $\Brbcl$    &       &  1.00    & -0.603   \\
 $\Brbcbl$   &       &          &  1.00    \\
\hline
\end{tabular}
%\caption{Correlation matrix of statistical uncertainties. 
%}
\end{center}

\subsection{Systematic uncertainties}

Since the $b$ reconstruction and the charge evaluation of the $b$-hadron 
 were done in the hemisphere where the lepton candidate was found, 
the correlation between the lepton selection  and the charge 
determination of the $b$ hadrons must be studied.
Although the lepton information was not included
in the training of the neural network to obtain the
charge of the $b$-hadron, a small correlation of
$\rho_{bl} = 1.036\pm0.005$ was found, where $\rho_{bl}$ 
represents the ratio of efficiencies to tag  a hemisphere
which contain a lepton over all hemispheres.  This was used
to reweight the Monte Carlo events, and twice the
statistical error on $\rho_{bl}$ was used to obtain the
contribution to the systematic uncertainty.

A more critical bias exists between the neural
network output and the $b$-hadron composition. 
The neural network output for a hemisphere containing a charged $b$-hadron 
was more likely to give the correct charge of the $b$-quark
than a hemisphere containing a neutral $b$-hadron.  
The effect of this bias was to increase the likelihood of 
incorrectly determining the charge of the $b$-quark for neutral
$b$-hadrons.  However, artificially adjusting the Monte
Carlo weight to account for this bias resulted in very
little change in the branching fractions.  A more critical
approach was to compare the measured branching fractions with the
ones obtained  without the charge separation.  
Without the separation, the lepton spectrum contained the
contributions from the direct decay and both modes of
the secondary decays.  The fit of the three modes was
performed by alternatively fixing one rate of
the two secondary decays modes, starting with the rate of 
$b\to\bar{c}\to\ell$ fixed to the result of the analysis,
until the fit converged.  
%Figure \ref{pic:fycomb} shows the distributions of the three modes as 
%a result of the fit. 
The difference between the branching ratios obtained in this fit
and the ones obtained with the charge separation
was used as a systematic uncertainty.

%\begin{figure}
%\begin{center}
%\epsfig{file=fycombbin.eps,width=120mm}
%\end{center}
%\caption{Results of fitting the lepton spectrum without
%the charge correlation.}
%\label{pic:fycomb}
%\end{figure}

The contributions to the systematic uncertainties 
of the correlation studies are shown in the first part of
Table \ref{Table:IV_sys}. Other sources considered for 
systematic uncertainties are
as follows:

\begin{itemize}
\item Lepton selection:\\
The muon and electron identification 
efficiencies and the background due to hadron misidentification 
were varied considering their measurement uncertainties in the data-simulation
comparisons (see Sections \ref{sec:muid}, \ref{sec:elecid}) as in Analysis I.
The residual contamination in the electron sample due to converted
photons has been varied by $\pm$ 10\%.

\item $b$-tagging\\
The efficiencies to tag $c$ and $uds$ quarks, as well as the
values of $R_b$ and $R_{uds}$, were varied in the same 
manner as in Analysis I. The correlation between the 
lifetime tag and the lepton tag was found to be 
$\rho_e=1.057\pm0.005$ and $\rho_\mu=1.041\pm0.005$.  These
values were varied by twice their statistical uncertainties.

\item Fitting \\
The uncertainty due to the finite Monte Carlo statistics in
the lepton spectrum fitting procedure was evaluated.

\item $b$-hadron composition\\
The production fraction for $\Lambda_b$ was taken
from \cite{ref:PDG} and set to
$(10.1^{+3.9}_{-3.1})\%$,
and the semileptonic branching fraction was
set to BR$(\Lambda_b\to\ell\nu~X)) = (7.4 \pm 1.1)\%$
\cite{ref:LAMBSL}.

\item Models\\
The mean fractional energy of $c$ hadrons
was  varied according to \cite{ref:LEPHF}.

The lepton distribution from the ``upper vertex'' was studied by
varying the contributions of $D_s \to \ell^-X$ and 
$\bar{D}^0(D^-) \to \ell^-X$ as suggested in reference \cite{ref:LEPHF}.

The modelling uncertainty related to the branching fractions assumed  
for $b\to\tau\to\ell$, $b\to J/\Psi\to\ell$ and to 
different lepton decay models was also calculated according to 
\cite{ref:PDG},\cite{ref:LEPHF} and \cite{ref:LEP99}. 

\end{itemize}

\begin{table}[ht]
\begin{tabular}{|l|c|c|c|c|}
\hline

Source & Range & $\Delta BR$ &$\Delta BR$
& $\Delta BR$\\
 & & $(b \to \ell)$ & $(b \to \bar{c} \to \ell)$ & $(b\to c \to \ell)$\\
 & & x$10^{-2}$ & x$10^{-2}$ & x$10^{-2}$\\
\hline
$\ell$-charge tag correlation & $\pm 1\%$   & $\mp0.08$ & $\mp0.03$  & $\mp0.09$ \\
NN bias on the $b$-charge   &    see text      & $\mp0.08$ & $\mp0.15 $ & $\mp0.11$ \\
$b$-hadron composition   &    see text      & $\mp0.04$ & $\mp0.02 $ & $\mp0.04$ \\
\hline

electron efficiency  & $\pm 3\%$         & $\mp0.18$ & $\mp0.04$ & $\mp0.15$ \\
muon efficiency& $\pm 2.5\%$       & $\mp0.13$ & $\mp0.05$ & $\mp0.10$ \\

Misidentified $e$   & $\pm 8\%$    & $\pm0.01$ & $\mp0.11$ & $\mp0.08$ \\
Misidentified $\mu$ & $\pm 6.5\%$  & $\pm0.01$ & $\mp0.08$ & $\mp0.05$ \\

Converted $\gamma$  & $\pm 10\%$   & $\pm0.01$ & $\mp0.04$ & $\mp0.03$ \\
%$P_b$   & $\pm 10\%$               & $\pm0.01$ & $\mp0.04$ & $\mp0.06$\\

\hline
%$b$-hadron species &   $\pm 5\%$   & $\mp0.04$ & $\mp0.02$ & $\mp0.03$\\
%$b$-tagging & $\pm 5\%$            & $\pm0.02$ & $\pm0.02$ & $\pm0.10$\\

$\varepsilon_c$     & $\pm 9 \%$   & $<$0.01    & $\mp$0.01 & $\mp$0.01 \\ 
$\varepsilon_{uds}$  & $\pm 22\%$  & $<$0.01    & $\pm$0.01 & $\mp$0.01 \\ 
$\ell$-b tag correlation & $\pm 1\%$   & $\mp$0.09  & $\mp$0.03 & $\mp$0.09 \\ 
$R_b$        & $0.21643\pm 0.00073$~\cite{ref:LEP99}
                         & $<$0.01    & $<$0.01   & $<$0.01  \\ 
$R_c$        & $0.1694\pm 0.0038$~\cite{ref:LEP99}
                         & $<$0.01    & $<$0.01   & $<$0.01 \\

\hline
MC statistics &                    & $\mp0.03$ & $\mp0.01$ & $\mp0.03$\\

\hline
$x_E(b)$     & $0.702 \pm 0.008$~\cite{ref:LEPHF}  & $\pm$0.03 & $\pm$0.05 & $\pm$0.07\\
$x_E(c)$     & $0.484 \pm 0.008$~\cite{ref:LEPHF}  & $\mp$0.01 & $\pm$0.01 & $\mp$0.01\\

${b\rightarrow W\rightarrow D}\over{b\rightarrow W\rightarrow D_s}$
 &$(1.28^{+1.52}_{-0.61})$~\cite{ref:LEPHF}  
&$^{+0.04}_{-0.04}$ &$^{-0.09}_{+0.08}$ & $^{+0.03}_{-0.03}$ \\

BR$(b\to\tau\to\ell)$ & $(0.459\pm0.071)\%$~\cite{ref:PDG}  
                      &$\mp0.02$&$\mp0.07$&$<0.01$\\
BR$(b\to J/\Psi\to\ell)$ &$(0.07\pm0.01)\%$~\cite{ref:PDG}  
                      &$\mp0.02$&$\pm0.01$&$\mp0.01$\\
BR$(c\to\ell)$ & $(9.85\pm0.32)\%$~\cite{ref:LEP99}  
                      & $\mp0.01$&$\mp0.05$&$\mp0.02$\\

\hline
\multicolumn{2}{|l|}{Total systematic}&$\pm0.28$&$\pm0.27$ & $\pm0.28$\\
\hline
\hline
\multicolumn{5}{|l|}{Decay models}\\
\hline
$b\to\ell$ model & ACCMM $(^{+ISGW}_{-ISGW**})$ &
        $^{-0.23}_{+0.42}$ & $^{+0.36}_{-0.58}$ & $^{+0.04}_{-0.04}$\\
$c\to\ell$ model & ACCMM1 $(^{+ACCMM2}_{-ACCMM3})$ &
        $^{-0.07}_{+0.07}$ & $^{+0.06}_{-0.05}$ & $^{-0.21}_{+0.09}$\\
\hline
\multicolumn{2}{|l|}{Total Models} & 
        $^{-0.24}_{+0.43}$ & $^{+0.36}_{-0.58}$ & $^{-0.21}_{+0.10}$\\
\hline
\end{tabular}
\caption{Analysis IV: Summary of systematic uncertainties.
Ranges given  in \% correspond to  relative variations around the
central value.}
\label{Table:IV_sys}
\end{table}

The summary of the different contributions to systematic uncertainties
 is given in Table \ref{Table:IV_sys}.
In conclusion, with the method of  charge correlation,
the following results have been obtained from the data collected with the 
DELPHI detector in 1994 and 1995:

\begin{eqnarray*}
BR(b \to \ell^-)            & = & (10.75 \pm 0.15 (stat) \pm 0.28 (syst)
 ^{-0.24}_{+0.43} (model))\%\\
BR(b \to c \to \ell^-)      & = & (7.99  \pm 0.27 (stat) \pm 0.28 (syst)
 ^{-0.21}_{+0.10} (model))\%\\
BR(b \to \bar{c}\to \ell^+) & = & (1.34  \pm 0.30 (stat) \pm 0.27 (syst)
 ^{+0.36}_{-0.58} (model))\%
\end{eqnarray*}

%**************************************************************************

\section {Combinations of results}
\label{sec:combi}

\begin{table}
\begin{center}
\begin{sideways}%
\begin{minipage}{\textheight}
\begin{center} 
\begin{tabular}{|l|c|c|c|c|}
\hline
           & Analysis I & Analysis II      & Analysis III      &Analysis IV \\
\hline
$\Brbl$\%  &{\boldmath$10.71\pm0.11\pm0.26^{-0.25}_{+0.42}$} &
            {\boldmath$10.78\pm0.14\pm0.28^{-0.34}_{+0.53}$} &
            {\boldmath$10.71\pm0.11\pm0.28^{-0.37}_{+0.44}$} &
            {\boldmath$10.75\pm0.15\pm0.28^{-0.24}_{+0.43}$}\\

$\Brbcl$\% &{\boldmath$ 8.05\pm0.39\pm0.38^{+0.23}_{-0.31}$} &
            {\boldmath$ 7.59\pm0.69\pm0.28^{-0.35}_{+0.50}$} &
                                                             &
            {\boldmath$ 7.99\pm0.27\pm0.28^{-0.21}_{+0.10}$}\\

$\Brbcbl$\%&{\boldmath$ 1.64\pm0.35\pm0.25^{+0.14}_{-0.23}$} &
            {\boldmath$ 2.00\pm0.49\pm0.27^{+0.56}_{-0.84}$} &
                                                             &
            {\boldmath$ 1.34\pm0.30\pm0.27^{+0.36}_{-0.58}$}\\
$( \Brbcl + $ & & & & \\
$\Brbcbl)\% $ & 9.69$\pm$0.24 $\pm$ 0.50$^{+0.37}_{-0.54}$ &
                9.59$\pm$0.30 $\pm$ 0.41$^{+0.29}_{-0.43}$ &
    {\boldmath$ 9.62 \pm 0.19  \pm  0.41^{+0.52}_{-0.49}$} &
                9.33$\pm$0.26 $\pm$ 0.52$^{+0.40}_{-0.64}$  \\
\hline
\end{tabular}
\end{center}
\caption[bla]{ Comparison of the results of the different analyses.
The measurements are shown using boldface characters, 
whereas slim-face  characters are used for sums which
are only shown for comparison.
The first uncertainty is statistical,
the second is systematic and the third is due to the  
uncertainty on the semileptonic model. 
}
\label{tab:allres}
\end{minipage}
\end{sideways}
\end{center}
\end{table}

A comparison of the results obtained in the different analyses described in 
the previous sections is shown in Table~\ref{tab:allres}.
A procedure to combine them in order to produce a final set of 
physical parameters has been developed. 
The basic technique, named Best Linear Unbiased Estimator 
(BLUE)~\cite{ref:blue}, 
determines the best estimate~$\hat{x}$ of a physical parameter built by 
a linear combination of measurements~$x_i$ obtained by several 
experiments; the coefficients of the combination 
are built from the covariance matrix~$\mbox{\sf E}_{ij}$ of the measured 
quantities. The method may be easily applied to determine several physical 
parameters simultaneously, by replacing that matrix with the more general 
one~$\mbox{\sf E}_{i\alpha j\beta}$ where the indices~$i,j$ refer to the 
experiments ( here analyses I to IV ) and~$\alpha,\beta$ identify the 
different physical parameters (here $\Brbl$~, $\Brbcl$ etc.). 

In order to apply this technique, it is necessary to estimate the 
full error matrix~{\sf E} including the off-diagonal elements; it has been 
determined as the sum of a statistical part and a systematic part with the 
latter accounting for the uncertainties on the parameters used by the 
analyses and obtained from other measurements. 

The statistical part has been built by splitting 
the statistical error~$\sigma_{i\alpha}$ of each 
parameter~$\alpha$ determined by the analysis~$i$ into two 
terms: the first one is computed from the observed number of leptons and is 
considered as fully correlated between different 
measurements; the other term is computed in order to keep invariant the 
total error and is assumed to be uncorrelated.

The estimation of the correlation between the parameters of different 
analyses is more 
complicated, as it is necessary to account for the correlation 
already present inside each single analysis. A reasonable criterion for that 
is to build the covariance elements by multiplying the correlated parts of
the two $\sigma_{i\alpha}$, 
described above, and by applying a correlation factor determined as an 
average of the correlation coefficients resulting from the different analyses. 

The described procedure can be applied only for identical data samples, while 
the different analyses used somewhat different data samples; as 
a consequence the full statistics has been divided into non-overlapping 
subsamples and the described procedure has been applied to each one of them. 
To do this the statistical 
uncertainties on the measurements have been scaled 
by the ratio of the square root of the number of events used by the 
corresponding analysis and the square root of the number of events in 
the subsample itself. 
These subsamples do not contain any common event and may be assumed 
uncorrelated; the total covariance matrix may then be obtained by summing the 
inverse of each covariance matrix and inverting again. 

A special care has been put in handling the results of the 
multivariate analysis which builds up the prompt  muon distributions by 
a linear combination of distributions obtained in 6~categories; 
the overlap with the $b$-tagged sample used by the other analyses has been 
conservatively assumed as corresponding to the category with the 
biggest purity and therefore the biggest weight. 

The systematic part of the error matrix has been evaluated by expressing 
a linear dependence on the external parameters of each result, and 
propagating the uncertainties on the parameters themselves; this corresponds 
to building up the sum of a set of error matrices, one for each uncertainty 
source, with correlation factors equal to~1 for all pairs of results 
affected by the corresponding external parameter, while the systematic 
errors relevant to only some of the results have been added as uncorrelated. 
The errors arising from the uncertainties on the decay models have not been 
used in the combination to obtain a result where the dependence on them is 
most explicit; as these errors give the biggest contribution to the total 
error this also protects from the instabilities described in the cited paper 
and in others dealing with this topic~\cite{ref:blue,ref:agost}~. 
The total systematic covariance matrix thus obtained has then been summed 
to the statistical covariance matrix; the inverse of the 
sum has been used to weight the four analyses results and find the combined 
value along with the total error. 

The following results have been obtained:
\begin{eqnarray*}
 \Brbl &=& (10.70\pm0.22 )\% \\
 \Brbcl&=& ( 7.98 \pm0.30 )\% \\
 \Brbcbl&=& (1.61\pm0.26 )\% \\
 \ci   &=&0.127  \pm0.014  \\
\end{eqnarray*}
where the total error, excluding model effect, is quoted; the 
global~$\chi^2$ of the fit is 1.52 for 12-4=8 degrees of freedom.

The statistical contribution to the total error has been obtained by 
propagating the statistical uncertainties on the four analyses output to 
the combined values. The systematic uncertainties 
breakdown on the combined values have been  
obtained by combining the error sets 
given for each analysis, using the same coefficients used to obtain the 
central values; this is equivalent to observing the effect of changing 
the combined values by~$1 \sigma$ for each of the error source. 
The full table of errors 
is shown in Table~\ref{tab:comb_sys}; 
the correlation matrix for the statistical 
and total uncertainties  is
shown in   Table~\ref{tab:comb_corr}. 

\begin{table}
\begin{center}
\begin{sideways}%
\begin{minipage}{\textheight}
\begin{center} 
\begin{tabular}{|l|c|c|c|c|c|}
\hline
Error Source &Range & $\Delta\Brbl$ &$\Delta\Brbcl$&$\Delta\Brbcbl$& $\Delta\ci$  \\ 
             &      & $10^{-2}$   & $10^{-2}$  & $10^{-2}$     & $10^{-2}$ \\ \hline

statistical         &             & $\mp$0.08  & $\mp$0.22 & $\mp$0.20& $\pm$1.3  \\ \hline
electron efficiency & $\pm 3 \%$  & $\mp$0.09  & $\mp$0.08 & $\mp$0.04& $\pm$0.01  \\ 
misidentified e      & $\pm 8 \%$ & $\mp$0.02  & $\mp$0.05 & $\mp$0.03& $\pm$0.04  \\ 
converted photons    & $\pm 10\%$ & $<$0.01    & $\mp$0.02 & $<$0.01& $\mp$0.03 \\ 
$\mu$ efficiency     &$\pm2.5\%$  &$\mp$0.15  & $\mp$0.12 & $\mp$0.04& $\mp$0.01  \\
misidentified $\mu$  &$\pm6.5\%;17\%$         
                     & $<$0.01   & $\mp$0.03  & $\mp$0.03 & $\mp$0.07  \\

\hline
 $\varepsilon_c$      & $\pm 9 \%$  & $\pm$0.01  & $\pm$0.01 &$\pm$0.03 &$\pm$0.02 \\ 
 $\varepsilon_{uds}$  & $\pm 22\%$  & $\pm$0.01  & $<$0.01 & $<$0.01 &$<$0.01 \\ 
 $\ell-b$ correlation & $\pm 1\%$  & $\mp$0.03  & $\mp$0.05 &$\mp$0.02 &$\mp$0.02 \\ 
\hline

other sources        &             & $\pm$0.09  & $\pm$0.10 &$\pm$0.05 &
$\pm0.5$    \\ 
\hline

$x_E(b)$      & $0.702 \pm 0.008$~\cite{ref:LEPHF}    & $\mp$0.01  & $\pm$0.03 &$\pm$0.02 &$\pm$0.05 \\
$x_E(c)$      & $0.484 \pm 0.008$   & $\mp$0.01  & $<$0.01 &$<$0.01 &$\pm$0.04 \\

${b\rightarrow W\rightarrow D}\over{b\rightarrow W\rightarrow D_s}$
 &$(1.28^{+1.52}_{-0.61})$~\cite{ref:LEPHF} 
 &$\pm$0.02 & $\pm$0.08 &$\mp$0.10 &$\mp$0.05 \\
BR($\btaul$)   &$(0.459\pm0.071)$\%~\cite{ref:PDG}  
                              & $\mp$0.01 & $\mp$0.02 &$\mp$0.08 &$\pm$0.04 \\
BR($\bpsill$)  &$(0.07\pm 0.01)$\%~\cite{ref:PDG}  
                              & $\mp$0.02 & $\mp$0.01 &$<$0.01 &$\mp$0.06 \\
BR($\cl$)      &$(9.85\pm0.32)$\%~\cite{ref:LEP99}  
                             & $\mp$0.01   & $<$0.01 &$\mp$0.02 &$\mp$0.01 \\ 
$\glcc$       & $(3.19\pm0.46)$\%~\cite{ref:LEP99}  
                             & $<$0.01   & $<$0.01   &$<$0.01  &$<$0.01\\
$\glbb$       & $(0.251\pm0.063)$\%~\cite{ref:LEP99}  
                             & $\mp$0.01   & $\mp$0.01   &$<$0.01&$\pm$0.01\\

\hline
 total systematic & & $\pm$0.21  & $\pm$0.21 &$\pm$0.17 &$\pm$0.5  \\ 
\hline
Semilept.mod.$\bl$\cite{ref:LEPHF}
              & ACCMM ($\mathrm{^{+ISGW}_{-ISGW**}}$)
&$^{-0.28}_{+0.44}$&$^{+0.10}_{-0.02}$&$^{+0.37}_{-0.47}$&$^{-0.3}_{+0.3}$\\

Semilept.mod.$\cl$\cite{ref:LEPHF}
              & ACCMM1($\mathrm{^{+ACCMM2}_{-ACCMM3}}$)  
              & $^{-0.09}_{+0.08}$ & $^{-0.19}_{+0.07}$& $^{+0.05}_{-0.04}$ &$^{-0.3}_{+0.3}$ \\ 

%\hline
%
% total models & &\large{$^{+0.44}_{-0.30}$} &\large{$^{+0.14}_{-0.20}$} &
%                 \large{$^{+0.31}_{-0.44}$} &\large{$\pm0.4$} \\ 

\hline
\end{tabular}
\end{center}
\caption[bla]{Systematic uncertainties associated to the 
combined results; the effect of sources relevant to only one analysis has 
been summarized in a single value labelled ``other sources''.}
\label{tab:comb_sys}
\end{minipage}
\end{sideways}
\end{center}
\end{table}

\begin{table}[htb]
%\begin{table}[p]
\begin{center}
\begin{tabular}{|l|cccc|}
\hline
          &  $\Brbl$ & $\Brbcl$ &$\Brbcbl$&  $\ci$ \\
\hline
 $\Brbl$     & 1.    & -0.066   & -0.051  &  0.018 \\
 $\Brbcl$    & 0.545 &  1.      & -0.733  & -0.091 \\
 $\Brbcbl$   & 0.231 & -0.277   &  1.     &  0.038 \\
 $\ci$       & 0.039 & -0.040   &  0.018  &  1.    \\
\hline
\end{tabular}
\end{center}
\caption{Correlation matrix of combined results. On the upper-right side 
the statistical coefficients are reported, on the lower-left side the 
statistical+systematic 
coefficients are shown.}
\label{tab:comb_corr} 
\end{table}

To investigate the effect of the main assumptions done in this combination
(~estimation of the correlated part of the error, estimation of the 
correlation coefficient between different parameters determined in different 
analyses~) the procedure has been repeated after changing them slightly.
The off-diagonal element in the error matrix has been 
changed using the most conservative assumption where a result does not add 
any information to another one having a smaller uncertainty. 
Different estimations of 
the correlation coefficient between different parameters in different 
analyses have also been tried. 
Compatible results have been obtained. 
The combination performed using a covariance matrix built from the 
statistical errors only was also found to give very similar results.

\section {Conclusions}

Four different analyses have been used to measure the
 semileptonic branching fractions for primary and cascade $b$ decays
%$\Brbl$, $\Brbcl$ and $\Brbcbl$
in hadronic \Zz decays from the data 
collected by the DELPHI experiment at LEP.
Results are compatible and a global average has been obtained: 
\par
\begin{eqnarray*}
 \Brbl &=& (10.70\pm0.08 (stat)\pm0.21( syst)_{+0.44}^{-0.30}(model) )\% \\
 \Brbcl&=& ( 7.98\pm0.22 (stat)\pm0.21( syst)^{+0.14}_{-0.20}(model) )\% \\
 \Brbcbl&=& (1.61\pm0.20 (stat)\pm0.17( syst)^{+0.30}_{-0.44}(model) )\% \\
 \ci   &=& 0.127\pm0.013 (stat)\pm0.005( syst)\pm0.004(model)\\
\end{eqnarray*}

The present result is compatible with and more precise than the previous 
DELPHI one \cite{ref:gam92}. It hence supersedes it.
It is also compatible with the recent results of the semileptonic branching 
fraction obtained at LEP \cite{Brnew} and with theoretical
calculations ~\cite{Neub}.

\newpage
%         Modified on 04-06-1999 by dimartino
%-------------------------------------------------------------------
\subsection*{Acknowledgements}
\vskip 3 mm
 We are greatly indebted to our technical 
collaborators, to the members of the CERN-SL Division for the excellent 
performance of the LEP collider, and to the funding agencies for their
support in building and operating the DELPHI detector.\\
We acknowledge in particular the support of \\
Austrian Federal Ministry of Science and Traffics, GZ 616.364/2-III/2a/98, \\
FNRS--FWO, Belgium,  \\
FINEP, CNPq, CAPES, FUJB and FAPERJ, Brazil, \\
Czech Ministry of Industry and Trade, GA CR 202/96/0450 and GA AVCR A1010521,\\
Danish Natural Research Council, \\
Commission of the European Communities (DG XII), \\
Direction des Sciences de la Mati$\grave{\mbox{\rm e}}$re, CEA, France, \\
Bundesministerium f$\ddot{\mbox{\rm u}}$r Bildung, Wissenschaft, Forschung 
und Technologie, Germany,\\
General Secretariat for Research and Technology, Greece, \\
National Science Foundation (NWO) and Foundation for Research on Matter (FOM),
The Netherlands, \\
Norwegian Research Council,  \\
State Committee for Scientific Research, Poland, 2P03B06015, 2P03B1116 and
SPUB/P03/178/98, \\
JNICT--Junta Nacional de Investiga\c{c}\~{a}o Cient\'{\i}fica 
e Tecnol$\acute{\mbox{\rm o}}$gica, Portugal, \\
Vedecka grantova agentura MS SR, Slovakia, Nr. 95/5195/134, \\
Ministry of Science and Technology of the Republic of Slovenia, \\
CICYT, Spain, AEN96--1661 and AEN96-1681,  \\
The Swedish Natural Science Research Council,      \\
Particle Physics and Astronomy Research Council, UK, \\
Department of Energy, USA, DE--FG02--94ER40817. \\
%=========================================================================%

\appendix
\section{Appendix}
\subsection{Single lepton likelihood}

The first part of the likelihood was constructed assuming a Poisson 
probability, using the single lepton spectra in data and simulation, 
subdivided in  $25 \times 25$ bins in the $(p_t,p_l)$ plane. 
The bins were chosen in such a way to have approximatively the same amount 
of data in each bins.
Nine classes were used, corresponding to the classes (a) to (g) mentioned in 
section \ref{sec:I_single}, with classes (f) and (g)  
splitted in two, for $b\bar b$ and non-$b\bar b$ events.  

$${\cal {L}}_1 = ln (L_1)  = \sum_{i=1}^{N_{bin}} \sum_{j=e,\mu} 
                     \{ DAT(i,j)ln (E(i,j)) - E(i,j) \}$$
$$E(i,j) = \sum_{\alpha=1}^{N_{class}} \{ {\cal P}(\alpha) MC(i,j,\alpha) \}$$
\noindent
where $DAT(i,j)$ represent the data and $MC(i,j)$ the simulated spectra, 
respectively. The ${\cal P}(\alpha) (\alpha=1,3)$ coefficients are the ratio 
between the unknown branching fractions and the corresponding values 
used in the simulation:

$${\cal P}(1) = \frac {\Brbl}{\Brbl _{sim}},
\ \ \ {\cal P}(2) = \frac {\Brbcl}{\Brbcl _{sim}},
\ \ \ {\cal P}(3) = \frac {\Brbcbl}{\Brbcbl _{sim}}$$
\noindent
whereas the ${\cal P}$ coefficients corresponding to lepton classes (d) to 
(g) are fixed to the values given in Table~\ref{tab:I_sys}.

\subsection{Di-lepton likelihood}

The second part of the likelihood was constructed assuming a Poissonian 
probability, using the di-lepton spectra in data and simulation, 
subdivided in  $7 \times 7$ bins in the combined momentum variables  
$(p_c^{min},p_c^{max})$.

The bins were chosen in such a way to have approximatively the same amount 
of data in each bins.
Twenty  classes were used,  
 according to the different possible combinations in the two opposite 
hemispheres  of the single-lepton classes (a) to (g) mentioned in 
section \ref{sec:I_single}.
\begin{eqnarray*}
{\cal{ L}}_2 = ln (L_2) = \sum_{i=1}^{M_{bin}} \sum_{j=ee,\mu\mu,e\mu} 
                 \{ &&DAT_{same}(i,j) ln (E_{same}(i,j)) - E_{same}(i,j) + \\ 
                    &&DAT_{opp.}(i,j) ln (E_{opp.}(i,j)) - E_{opp.}(i,j) \}
\end{eqnarray*}
$$E_{same}(i,j) = \sum_{\alpha=1}^{M_{class}} \{ {\cal S}(\alpha) 
                                                  MC_{same}(i,j,\alpha) \}$$
$$E_{opp.}(i,j) = \sum_{\alpha=1}^{M_{class}} \{ {\cal O}(\alpha) 
                                                  MC_{opp.}(i,j,\alpha) \}$$
\noindent
where $DAT_{same}(i,j)$ ( $DAT_{opp.}(i,j)$ ) represent the spectra of 
di-leptons in data, in opposite hemispheres, having the same (opposite) 
charge and $MC_{same}(i,j)$ ($MC_{opp.}(i,j)$) represent the 
simulated spectra. 
The ${\cal S}(\alpha)$ ( ${\cal O}(\alpha)$ ) coefficients depend on the 
ratio between 
the unknown branching fractions and the corresponding values used in the 
simulation and on the mixing probability $\ci$.
For example for the first and the second classes, containing 
($\btol,\btol$) and ($\btol,\bctol$)  di-leptons, respectively:
\begin{eqnarray*}
{\cal S}(1) &=& 2\ci (1-\ci) {\cal P}(1)^2 = 
                2\ci (1-\ci) ( \frac{\Brbl} {\Brbl _{sim}} )^2 \\
{\cal O}(1) &=& ( 1- 2\ci (1-\ci) ) {\cal P}(1)^2 = 
                ( 1- 2\ci (1-\ci) ) ( \frac{\Brbl} {\Brbl _{sim}} )^2 \\
{\cal S}(2) &=& ( 1 - 2\ci (1-\ci) ) {\cal P}(1) {\cal P}(2) = 
       ( 1 - 2\ci (1-\ci) ) \frac{\Brbl\Brbcl} {\Brbl_{sim} \Brbcl_{sim} } \\
{\cal O}(2) &=& 2\ci (1-\ci) {\cal P}(1) {\cal P}(2) = 
                2\ci (1-\ci) \frac{ \Brbl\Brbcl} {\Brbl_{sim} \Brbcl_{sim} } 
\end{eqnarray*}
The total likelihood is the sum of the single and the di-lepton likelihoods:

$${\cal{L = L}}_1 + {\cal{L}}_2$$ 

In the fit $P(1),P(2)$,$P(3)$ and $\ci$ are free parameters, whereas the 
${\cal P}$ coefficients corresponding to lepton classes (d) to (g) are fixed 
to the values given in Table~\ref{tab:I_sys}.

\end{document}